\documentclass[amsmath,amssymb,aps,twocolumn]{revtex4-2}
\usepackage{graphicx,amsmath,amsfonts,bm, color,MnSymbol}
\usepackage{graphicx}
\usepackage{bm}
\usepackage{xcolor,hyperref}
\hypersetup{
   colorlinks,
   linkcolor={blue!50!black},
   citecolor={blue!50!black},
   urlcolor={blue!80!black}
}
\newcommand*\colvec[3][]{\begin{pmatrix}\ifx\relax#1\relax\else#1\\\fi#2\\#3\end{pmatrix}}
\newcommand*\rowvec[3][]{\begin{pmatrix}\ifx\relax#1\relax\else#1&\fi#2&#3\end{pmatrix}}

\newcommand{\mycomment}[1]{}

\newcommand{\LB}{\Delta_{\text{LB}}}
\newcommand{\bra}[1]{\left.\langle #1 \right.|}
\newcommand{\ket}[1]{\left.| #1 \right.\rangle}
\newcommand{\braket}[2]{\langle #1 | #2 \rangle}

\begin{document}

\title{Deformation-induced spectral softening drives phase separation on non-spherical membranes}
\author{Avraham Moriel and Howard A. Stone}
\affiliation{Department of Mechanical and Aerospace Engineering, Princeton University, Princeton, New Jersey 08544, USA}

\begin{abstract}
Models of curvature-driven membrane sorting couple composition directly to membrane bending elasticity. Here we show that shape deformations can destabilize a homogeneous state without this composition-dependent mechanical coupling. We analyze the eigenspectrum of the Laplace-Beltrami operator on deformed spheres analytically, and verify it numerically. As the deformation amplitude increases, the spectral degeneracy breaks, and the lowest nonzero eigenvalue decreases substantially. Using Cahn-Hilliard dynamics, we show that this eigenvalue softening provides a purely geometric route to phase separation.
\end{abstract}
\maketitle

Biological membranes serve as underlying two-dimensional manifolds for lateral diffusion --- a transport mechanism essential for cellular signaling~\cite{lindblom2009lipid,ALMEIDA1995305,kusumi2012dynamic,axelrod1983lateral,SAXTON1999229}. The transport dynamics along this heterogeneous ``fluid mosaic''~\cite{singer1971molecular,singer1972fluid,singer1974molecular,lingwood2010lipid,engelman2005membranes} are governed by hydrodynamic interactions~\cite{peters1982lateral,gambin2006lateral,saffman1975brownian}, and importantly, are tightly coupled to the membrane's underlying geometry~\cite{henle2010hydrodynamics,phillips2009emerging,baumgart2003imaging,bhatia2009amphipathic,hatzakis2009curved,parthasarathy2006curvaturereview,heinrich2010dynamic}. Diffusing constituents can alter the membrane mechanics either via a preferred intrinsic curvature, or a modified bending stiffness~\cite{Zimmerberg2006,kawakatsu1993phase,seifert1993viscous,bassereau2018,netz1995inhomogeneous,helfrich1973elastic,winter2025phase,julicher1996shape,julicher1993domain,leibler1986curvature,seifert1993curvature}, resulting in a dynamic composition-geometry feedback --- providing a fertile framework for understanding domain formation, budding, and other morphological transitions~\cite{leibler1986curvature,seifert1993curvature,seifert1997configurations,julicher1996shape,julicher1993domain,reister2010diffusing,winter2025phase,huang2006curvature,mukhopadhyay2008lipid,jiang2008curvature,jiang2012dynamic,reynwar2007aggregation,aimon2014membrane,roux2010membrane,sorre2012nature,beltran2019membrane}. This framework relies on an explicit mechanical coupling between the constituent's concentration and the membrane's curvature, and assumes the membrane is free to deform dynamically.

Recent observations indicate that mechanical coupling may play a lesser role in some cases~\cite{tian2009sorting,kamal2009measurement}. Micropipette-aspiration experiments demonstrated strong curvature-dependent lipid sorting in membrane nanotubes near a demixing point~\cite{sorre2009curvature}. Experiments utilizing solid-supported lipid bilayers on patterned substrates~\cite{parthasarathy2006curvature,yoon2006topographic,hsieh2012curvature} and rigid nanoparticle scaffolds~\cite{black2014membrane} showed that an underlying static geometry alone is sufficient to induce or inhibit lateral phase separation, independent of membrane deformability. This purely geometric sorting is also observed \emph{in vivo} --- cardiolipin spontaneously diffuses and localizes to the highly-curved, rigidly-fixed, poles of bacterial membranes~\cite{renner2011cardiolipin}. Many other biological processes also exhibit temporally persistent, highly-deformed geometries~\cite{salbreux2012actin} (e.g., cytokinesis, endocytosis, and gastrulation). Because the timescales associated with such processes are much longer than those of microscopic diffusion, lateral transport could be affected by the underlying non-isotropic geometries. What drives phase separation under weak mechanical coupling, when the membrane geometry is fixed?

\begin{figure}[b]
\includegraphics[width=0.45\textwidth]{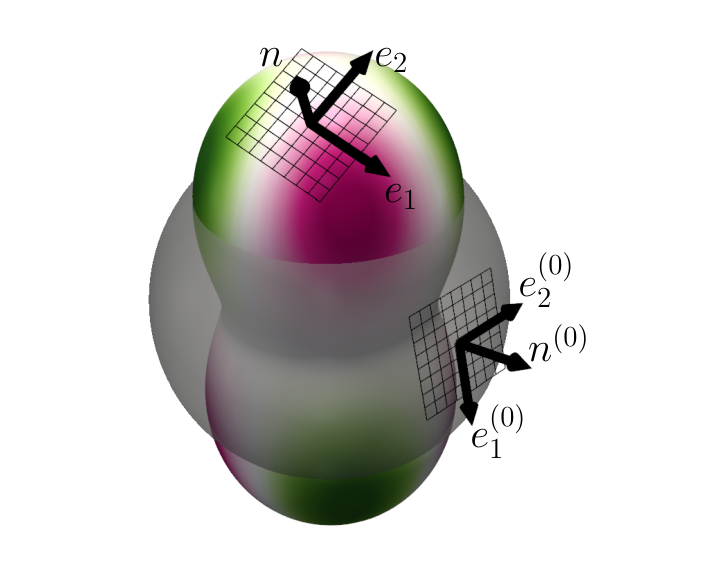}
\caption{Undeformed and deformed surfaces. The undeformed sphere in gray, with the local tangents $\bm{e}_1^{(0)}$ and $\bm{e}_2^{(0)}$, and the normal $\bm{n}^{(0)}$. The deformed sphere (in color) with the local tangents $\bm{e}_1$ and $\bm{e}_2$, and the normal $\bm{n}$. The green and pink colors schematically indicate a scalar density field transported on the deformed sphere.}
\label{fig:fig1}
\end{figure}

A purely geometric mechanism may explain this phenomenon. We demonstrate how geometry affects single-component diffusion dynamics and triggers phase separation without any explicit coupling to the underlying curvature, and assuming that the diffusivity is spatially homogeneous~\cite{saffman1975brownian,henle2010hydrodynamics}. First, we extend a previously-used perturbative approach~\cite{frank2019pinning} to examine how small deformations affect the eigenspectrum of the Laplace-Beltrami operator of a sphere. We demonstrate numerically that the minimal eigenvalue decreases as the deformation amplitude increases. Consequently, we simulate Cahn-Hilliard nonlinear surface dynamics for various shape perturbations, showing that deformations eventually induce phase separation.

\emph{Laplace-Beltrami spectrum on a deformed sphere.} --- We consider a reference surface $\Omega^{(0)}$, characterized by the local (covariant) tangent vectors $\bm{e}^{(0)}_a$ (for $a\!=\!1,2$), the unit normal $\bm{n}^{(0)}\!\equiv\!\frac{\bm{e}_1^{(0)}\times\bm{e}_2^{(0)}}{|\bm{e}_1^{(0)}\times\bm{e}_2^{(0)}|}$, the metric tensor $g_{ab}^{(0)}\!\equiv\!\bm{e}_a^{(0)}\cdot\bm{e}_b^{(0)}$, and $g^{(0)}\equiv\det \bm{g}^{(0)}$ (where the tensor $\bm{g}^{(0)}$ has the components $g_{ab}^{(0)}$), depicted in Fig.~\ref{fig:fig1} (in gray).

\begin{figure}[t!]
\includegraphics[width=0.48\textwidth]{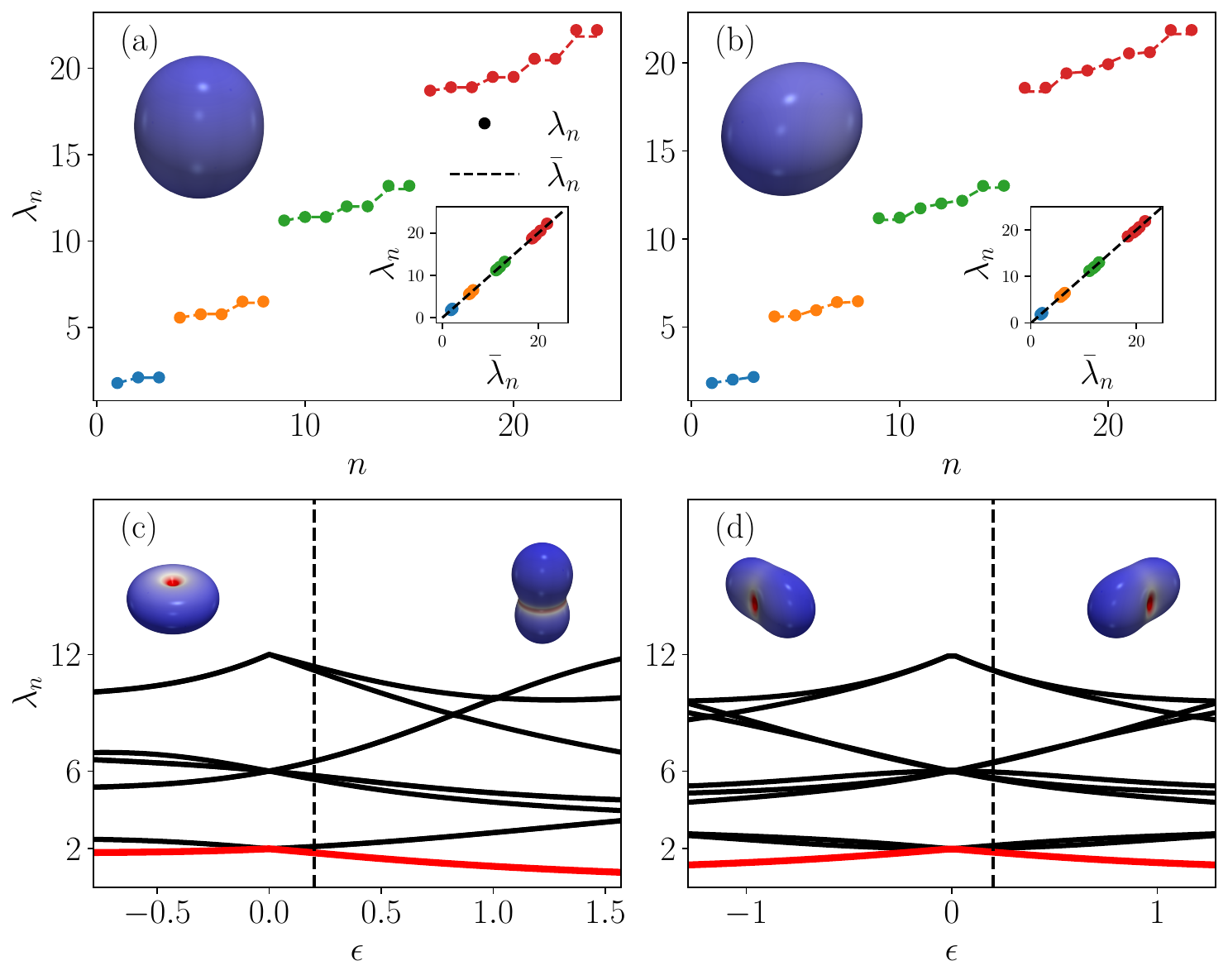}
\caption{Laplace-Beltrami spectra on deformed spheres. (a) The numerically computed eigenvalues $\lambda_n\!\equiv\!-\Lambda_n$ (points) and the corresponding analytical eigenvalues $\bar{\lambda}_n$ obtained to leading order in $\delta h$ (dashed lines), for the first four spectral bands of a deformed sphere with $\delta h \!=\! 0.2 Y_{2,0}$. Top inset: the deformed sphere (colors represent the local Gaussian curvature). Bottom inset: the numerical $\lambda_n$ versus the analytical $\bar{\lambda}_n$ eigenvalues, with the diagonal $\lambda_n\!=\!\bar\lambda_n$ as a guide-to-the-eye. (b) A similar plot for $\delta h\!=\!0.2 \left(\frac{Y_{2,2} + Y_{2,-2}}{\sqrt{2}}\right)$. (c) The numerically computed 10 lowest non-zero eigenvalues for a deformed sphere with $\delta h\!=\!\epsilon Y_{2,0}$, as a function of the deformation amplitude $\epsilon$. The lowest non-zero eigenvalue is annotated as a red thick line. The vertical dashed line corresponds to the spectrum shown in (a). Insets: The resulting deformed oblate (left) and prolate (right) shapes with $\epsilon\!=\!-1$ and $\epsilon\!=\!1$ respectively. (d) A similar plot for the perturbation $\delta h\!=\!\epsilon \left(\frac{Y_{2,2} + Y_{2,-2}}{\sqrt{2}}\right)$. Insets: The deformed spheres with $\epsilon\!=\!-1$ and $\epsilon\!=\!1$ (left and right respectively).}
\label{fig:fig2}
\end{figure}

A scalar density field $\phi$ can (linearly) diffuse on $\Omega^{(0)}$, evolving according to $\partial_t \phi \!=\!D_s\LB^{(0)} \phi$. Particle diffusion on membranes typically renders $D_s$ geometrically-dependent~\cite{saffman1975brownian,henle2010hydrodynamics}, yet here we set it to a fixed surface diffusion coefficient ($D_s\!=\!1$ in what follows), and $\LB^{(0)}$ is the (scalar) Laplace-Beltrami operator on the surface $\Omega^{(0)}$, cast as~\cite{bowick2009two,deserno2015fluid}
\begin{equation}\label{eq:LB0}
\LB^{(0)}\bullet\equiv\!\frac{1}{\sqrt{g^{(0)}}}\partial_a\left(\sqrt{g^{(0)}} g^{ab}_{(0)} \partial_b\bullet\right) \ ,
\end{equation}
using Einstein summation over repeated indices. The operator $\LB^{(0)}$ has a spectrum of eigenvalues $\Lambda^{(0)}_n$, and associated eigenmodes $\psi_n^{(0)}$ (with $n$ as an index), obeying $\LB^{(0)}\ket{\psi_n^{(0)}}\!=\!\Lambda^{(0)}_n\ket{\psi_n^{(0)}}$, where $\psi_n^{(0)}$ are normalized to unity under the metric $\bm{g}^{(0)}$, as $\braket{\psi_n^{(0)}}{\psi_m^{(0)}}\!=\!\delta_{nm}$ ($\delta_{nm}$ being Kronecker delta function). When the reference shape is the unit sphere, $\Omega^{(0)}\!=\!\mathbb{S}^2$, a solution to the diffusion problem can be expressed as a weighted sum of the eigenmodes of $\LB^{(0)}$ --- the spherical harmonics $Y_{l,m}$, where $l$ and $m$ serve as indices instead of the index $n$ --- evolving temporally as $\exp\left[\Lambda_l t\right]$ with their associated eigenvalues $\Lambda_l\!=\!-l(l+1)$. 

To understand the role of geometry on diffusion, we impose a small, spatially-dependent, surface perturbation in the normal direction, $\delta h(\bm{u}) \bm{n}^{(0)}$ ($\bm{u}$ are the surface coordinates). A similar perturbative approach combined with conformal mapping was used in~\cite{frank2019pinning} to obtain the leading-order corrections on cylinders and spheres. The deformed surface $\Omega^{(1)}$, depicted schematically in Fig.~\ref{fig:fig1} (in color), produces a different $\LB^{(1)}$ operator. For small deformations, we expand $\LB^{(1)}$ retaining leading-order terms in $\delta h$~\cite{SM}, yielding  \begin{equation}\label{eq:LB1}
\begin{split}\LB^{(1)} \bullet\simeq 
     &\left(1-K^{(0)} \delta h\right)\LB^{(0)}\bullet + \\ &\frac{1}{\sqrt{g^{(0)}}}\partial_{a}\left[\sqrt{g^{(0)}} \left( K^{(0)} g^{ab}_{(0)}  - 2 K^{ab}_{(0)}\right)\delta h\partial_{b}\bullet\right] \ ,
        \end{split}
\end{equation}
where $K_{ab}^{(0)}\!\equiv\!\bm{e}_a^{(0)}\cdot \partial_b \bm{n}^{(0)}$ is the extrinsic curvature tensor (the second fundamental form), $K^{ab}_{(0)}\!=\!g^{ac}_{(0)}g^{bd}_{(0)}K_{cd}^{(0)}$, and $K^{(0)}\!=\!g_{(0)}^{ab}K_{ab}^{(0)}$ is the total extrinsic curvature of the reference surface~\cite{jenkins1977equations,jenkins1977static,zhong1987instability,zhong1989bending,capovilla2003deformations,spivak1970comprehensive,deserno2015fluid,SM}.
Using $\Omega^{(0)}\!=\!\mathbb{S}^2$, $g_{(0)}^{ab}\!=\!\text{diag}\left(1,\sin^{-2}\theta\right)\!=\!K^{ab}_{(0)}$, and $K^{(0)}\!=\!2$, yielding $\LB^{(1)}\!\simeq\!\left(1-2\delta h\right)\LB^{(0)}$, indicating that the imposed normal perturbation changes the emerging spectrum, and consequently the diffusive dynamics.

Spherical shape deformations can be expanded using spherical harmonics. We can then use degenerate perturbation theory to obtain the leading-order corrections to the eigenvalues by using the unperturbed bases $\psi_n^{(0)}$ of $\LB^{(0)}$. Expressing $\LB^{(1)}\!\simeq\! \LB^{(0)} + \delta \LB$, and $\Lambda^{(1)}_n\!\simeq\!\Lambda^{(0)}_n+\delta\Lambda_n$, and considering an initially-degenerate spectrum (e.g., for a sphere), we write $\psi_{n}\!\simeq\!\sum_i \alpha_i\psi_{n,i}^{(0)}+\delta \psi_{n}$, where $n$ indicates the degenerate band, $i$ is the index within the band (equivalent to $l$ and $m$ for $Y_{l,m}$, respectively), and $\alpha_i$ are coefficients to be determined. Expanding the eigenvalue equation and contracting with $\bra{\psi_{n,j}^{(0)}}$, we find that for a sphere $\delta\Lambda_{n}$ are the eigenvalues of $2l(l+1)\bra{Y_{l,m_1}}\delta h\ket{Y_{l,m_2}}$, governed by the spatial overlap of the surface perturbation $\delta h$ with the eigenmode basis functions $Y_{l,m_1}$ and $Y_{l,m_2}$~\cite{SM}.

We show in Fig.~\ref{fig:fig2} two examples of the emerging spectra of deformed spheres under different spatial perturbations, where for convenience we define $\lambda_n\!\equiv\!-\Lambda_n$. In each case we first deform a sphere with $\delta h(\bm{u}) \bm{n}^{(0)}$, and then rescale the surfaces to $4\pi$, conserving the surface area across deformations. Expressing $\delta h$ using spherical harmonics $Y_{l_p,m_p}$, we use $\delta h\!=\! 0.2 Y_{2,0}$ ($l_p\!=\!2$, and $m_p\!=\!0$) to perturb the sphere and diagonalize the modified $\LB$ operator numerically, leading to the spectrum $\lambda_n$ presented as points in panel Fig.~\ref{fig:fig2}(a). Our analytical calculation~\cite{SM} leads to $\bar{\lambda}_n$ presented as dashed lines, agreeing with the numerical results. To further demonstrate this agreement we plot the inset $\lambda_n$ versus $\bar{\lambda}_n$, with the diagonal line $\lambda_n\!=\!\bar{\lambda}_n$ as a guide-to-the-eye. 

In panel (b) we use $\delta h\!=\!0.2 \left(\frac{Y_{2,2} + Y_{2,-2}}{\sqrt{2}}\right)$, demonstrating that the symmetry of $\delta h$ affects the breaking of the degeneracy and the redistribution of eigenvalues within the bands. The perturbative expansion holds well also for different $l_p,m_p$ combinations, allowing us to deduce, for example, that the linear-order correction vanishes for all $\delta h$ with odd $l_p$ values, as a consequence of Wigner selection rules~\cite{SM}. This analytical step prepares the stage for larger amplitude perturbations.

Next, we probe the emerging spectra under large deformation amplitudes, mimicking irregular biophysical shapes. Considering $\delta h\!=\!\epsilon Y_{2,0}$, we numerically compute the eigenvalue spectrum for increasing values of the deformation amplitude $\epsilon$. We plot the 10 lowest non-zero eigenvalues as a function of $\epsilon$ in Fig.~\ref{fig:fig2}(c). Clearly, the emerging spectra are not symmetric with $\epsilon$: $\epsilon\!>\!0$ yields a prolate shape, elongating a single axis while compressing the other two directions, while $\epsilon\!<\!0$ produces an oblate spheroid, shown by the insets. In both cases, as $|\epsilon|$ increases, the degeneracy breaks, and the eigenvalue bands exhibit level crossings. We repeat a similar process with $\delta h\!=\!\epsilon\left(\frac{Y_{2,2}+Y_{2,-2}}{\sqrt{2}}\right)$ and show the resulting eigenvalues in panel (d). In this case, the emerging spectra are symmetric about $\epsilon\!=\!0$: the perturbation $Y_{2,2}+Y_{2,-2}$ yields similar shapes irrespective of the sign of $\epsilon$, enforcing the spectral symmetry around $\epsilon\!=\!0$. 

Generically, deformations break the spectral degeneracy, implying that lateral diffusion becomes anisotropic, with relaxation times that crucially depend on the spatial symmetry of the modes with respect to the overall deformed shape. Additionally, Fig.~\ref{fig:fig2}(c)-(d) show that the lowest non-zero eigenvalue $\lambda_{\text{min}}$ decreases as the sphere deforms. This ``softening'' of the lowest eigenvalue has a general geometric origin known as the Hersch's inequality ~\cite{hersch1970quatre}), implying that global diffusive processes take more time on deformed spheres. 

\emph{Deformation-induced phase separation. ---} The decrease in $\lambda_{\text{min}}$ gives rise to anisotropic global relaxation (despite being locally isotropic), but could also destabilize the homogeneous state. Recent works explored the role of geometry on reaction-diffusion dynamics, demonstrating how shape perturbations induce nontrivial dynamics (e.g., pattern pinning or modified wave propagation)~\cite{frank2019pinning,nishide2025oscillatory,nishide2022pattern,tamemoto2021reaction,noguchi2023disappearance,singh2022sensing,orlandini2013domain}. Here we utilize a Landau-Ginzburg $\phi^4$ free energy functional~\cite{ginzburg2009theory,kardar2007statistical,safran2018statistical} for the scalar concentration field $\phi$, \begin{equation} \label{eq:LG}
    F=\int \left[\frac{1}{4}\left(\phi^2 - 1\right)^2 + \frac{\eta}{2}\left(\nabla \phi\right)^2 + F_{\text{m}}\right]  d\Omega\,
\end{equation}
representing a mixture that locally favors demixing, $\eta$ is a phenomenological constant, $F_\text{m}$ represents the mechanical free energy density, and $\int \bullet \ d\Omega$ indicates integration over the surface~\cite{SM}. This functional is subjected to the mass conservation on the surface, expressed here using the average concentration $\langle\phi\rangle\!\equiv\!\frac{1}{|\Omega|}\int\phi d\Omega\!=\!\phi_0$ ($|\Omega|$ denotes the total surface area). Past works have explored the coupling between the mechanical free energy density~\cite{helfrich1973elastic,seifert1993curvature,julicher1993domain,julicher1996shape,winter2025phase,leibler1986curvature}, e.g., $F_{\text{m}} \!=\! \frac{\kappa(\phi)}{2}\left[H-H_0(\phi)\right]^2$ (where $\kappa(\phi)$ and $H_0(\phi)$ are the concentration-dependent bending modulus and spontaneous curvature respectively) and the emerging phase-separation process --- here we explicitly set $F_\text{m}\!=\!0$, depicting the case where such contributions are negligible with respect to the mixing terms (and as a result, the underlying geometry is fixed). 

Expanding $\phi$ in $\psi_n$s, modes with $\eta \lambda_n + 3 \phi_0^2 \!>\!1$ are linearly stable~\cite{cahn1961spinodal}. Because $\phi$ is conserved on $\Omega$, spatially-uniform fluctuations are forbidden, implying that the lowest non-zero eigenmode is the most prone to instability. To demonstrate the role of geometry, we first fix $\eta\!=\!1\!>\!\eta_c$ and $\phi_0\!=\!0$, and focus on $\delta h$ with $l_p\!=\!2$ and $m_p\!=\!0$, as these resemble cell division geometries (the dynamics demonstrated below occurs in a similar way for other spatial perturbations~\cite{SM}). As $\epsilon$ increases, $\lambda_{\text{min}}$ decreases, until $\lambda_{\text{min}}\!<\!1$, allowing for a deformation-induced phase separation to ensue [see Fig.~\ref{fig:fig2}(c)]. The same behavior could also be explained by viewing the energetic costs of bulk phases and interfaces --- the geometrical deformation allows for shorter interfaces to form, facilitating phase separation~\cite{SM}.

\begin{figure}[b!]
\includegraphics[width=0.48\textwidth]{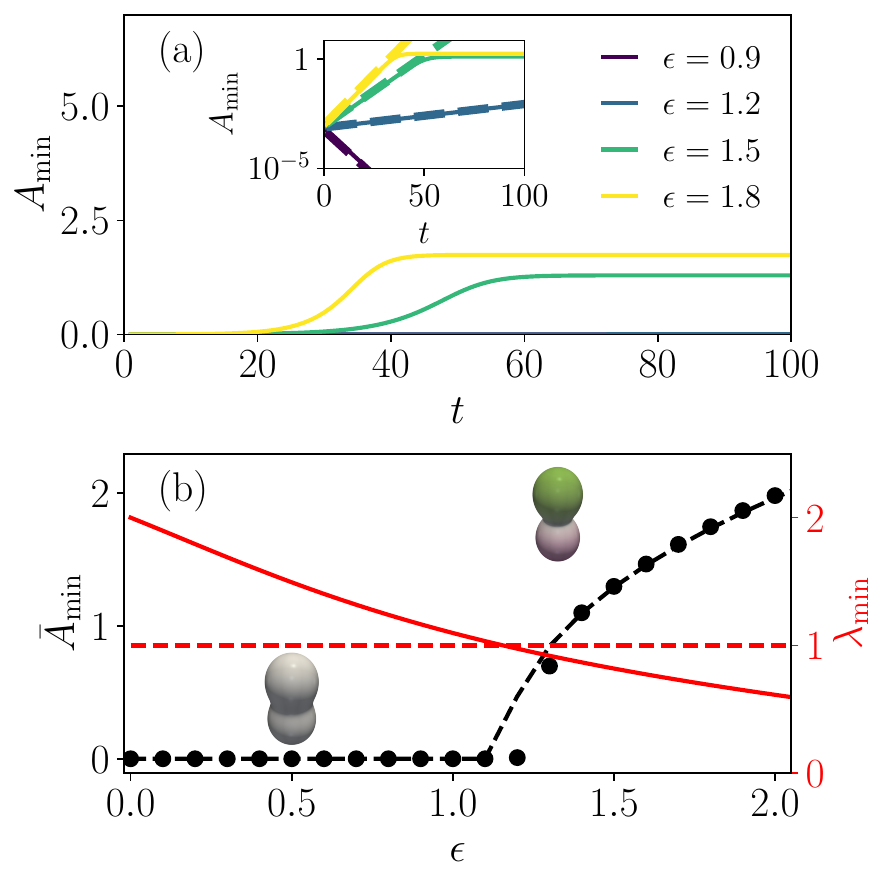}
\caption{Deformation-induced phase separation. (a) $A_{\text{min}}(t)\!\equiv\!\braket{\psi_{\text{min}}}{\phi(\bm{u},t)}$ dynamics obtained for  $\delta h\!=\!\epsilon Y_{2,0}$ for different $\epsilon$ values. Inset: the dynamics in semi-log plot, showing an instability growth rate of $\lambda_\text{min}(1-\lambda_\text{min})$ in dashed lines. (b) The steady-state values  $\bar{A}_{\text{min}}$ as a function of $\epsilon$ (in black circles, left axis), showing a transition from $\bar{A}_{\text{min}}\!=\!0$ to $\bar{A}_{\text{min}}\!\neq\!0$ around $\epsilon_c\!\simeq\!1.3$. We also plot the theoretical prediction $\bar{A}_{\text{min}}\!\sim\!\sqrt{1-\lambda_{\text{min}}}$ for $\epsilon\!>\!\epsilon_c$~\cite{SM} (dashed black line), and $\lambda_{\text{min}}$ in red (right axis), to demonstrate that the transition occurs once $\lambda_{\text{min}}\!\simeq\!1$. Left and right insets show a mixed, final $\phi$ distribution on a deformed sphere with $\epsilon\!=\!1$, and a phase-separated distribution for $\epsilon\!=\!1.5$ respectively.}
\label{fig:fig3}
\end{figure}

We demonstrate this effect numerically by solving the Cahn-Hilliard~\cite{cahn1958interfacial} equation emerging from Eq.~\eqref{eq:LG} on deformed spheres~\cite{SM}, $\partial_t \phi\!=\!\nabla^2\mu$ with $\mu\!=\!\frac{\delta F}{\delta \phi}$. We initiate a random concentration field $\phi$ (obeying $\langle\phi\rangle\!=\!\phi_0$) on the deformed sphere, and simulate the Cahn-Hilliard dynamics stemming from Eq.~\eqref{eq:LG} (conserving $\langle\phi\rangle$ by construction). This yields the concentration dynamics over time $\phi(\bm{u},t)$. We project the emerging solutions on $\psi_{\text{min}}$ --- the eigenmode corresponding to the minimal eigenvalue $\lambda_{\text{min}}$ --- defining the projection amplitude $A_{\text{min}}(t)\!\equiv\!\braket{\psi_{\text{min}}}{\phi(\bm{u},t)}$. If the mixed state prevails, $A_{\text{min}}(t)\!\rightarrow\!0$ at long times. If $A_{\text{min}}(t)$ converges to a finite value, the obtained spatial profile is heterogeneous, approaching a spatial form similar to $\psi_{\text{min}}$.


We show in Fig.~\ref{fig:fig3}(a) the $A_{\text{min}}(t)$ dynamics obtained for various values of $\epsilon$. As anticipated, above a critical $\epsilon$ value, $\epsilon_c$, $A_{\text{min}}$ grows in time as $\exp\left[\lambda_{\text{min}}(1-\lambda_{\text{min}})\right]$ [see the inset of Fig.~\ref{fig:fig3}(a)] until it saturates to its steady-state value $\bar{A}_{\text{min}}$~\cite{SM}. We show in Fig.~\ref{fig:fig3}(b) that $\bar{A}_{\text{min}}$ transitions from $\bar{A}_{\text{min}}\!\rightarrow\!0$ for $\epsilon\!<\!\epsilon_c$, to $\bar{A}_{\text{min}}\!\sim\!\sqrt{1-\lambda_{\text{min}}}$ for $\epsilon\!>\!\epsilon_c$~\cite{SM}, plotted as a dashed black line. As a guide-to-the-eye, we also plot $\lambda_{\text{min}}(\epsilon)$ in red, to demonstrate this transition occurs once $\lambda_{\text{min}}\!=\!1$. 

\begin{figure}[t]
\includegraphics[width=0.42\textwidth]{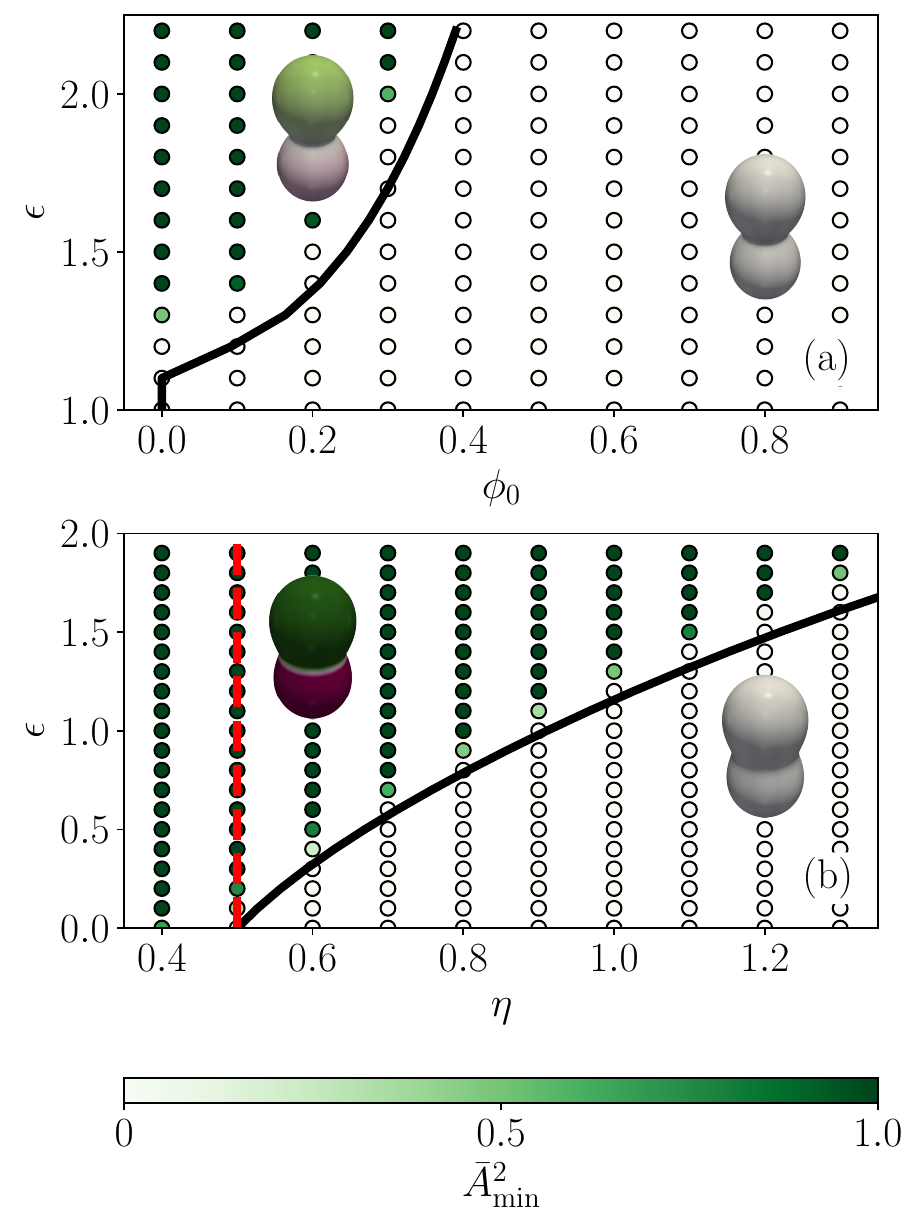}
\caption{Phase diagrams of the deformation-induced instability. (a) The $\phi_0\!-\!\epsilon$ phase diagram (using $\eta\!=\!1$), demonstrating phase separation occurs for high $\epsilon$ values for various values of $\phi_0$ (the prediction $\phi_c\!=\!\sqrt{\frac{1 - \eta \lambda_{\text{min}}}{3}}$ for $\eta\!=\!1$ is plotted as a black curve~\cite{SM}). Insets demonstrate the phase-separated ($\phi_0\!=\!0.2$, $\epsilon\!=\!1.5$, top-left) and mixed ($\phi_0\!=\!0.6$, $\epsilon\!=\!1.5$, bottom-right) regimes. (b) The $\eta\!-\!\epsilon$ phase diagram (using $\phi_0\!=\!0$), showing that above $\eta\!=\!\frac{1}{2}$ the instability occurs only for sufficiently large $\epsilon$ values (the theoretical prediction $\eta_c\!=\!\lambda_{\text{min}}^{-1}$ is plotted as a black curve~\cite{SM}). Below $\eta\!=\!\frac{1}{2}$ the lowest mode is already unstable for the undeformed sphere (dashed, vertical red line). Insets demonstrate the phase-separated ($\eta\!=\!0.6$,  $\epsilon\!=\!1$, top-left) and mixed ($\eta\!=\!1.1$,  $\epsilon\!=\!1$, bottom-right) regimes.}
\label{fig:fig4}
\end{figure}

We used the numerical simulations to probe the emerging phase diagrams for different $\phi_0$ and $\eta$ values. We show the $\phi_0-\epsilon$
phase diagram in Fig.~\ref{fig:fig4}(a) (for $\eta\!=\!1$), and the $\eta-\epsilon$ phase diagram (for $\phi_0\!=\!0$) in Fig.~\ref{fig:fig4}(b). The plotted theoretical predictions agree with the numerically-obtained phase diagrams.

\begin{figure}[t]
\includegraphics[width=0.42\textwidth]{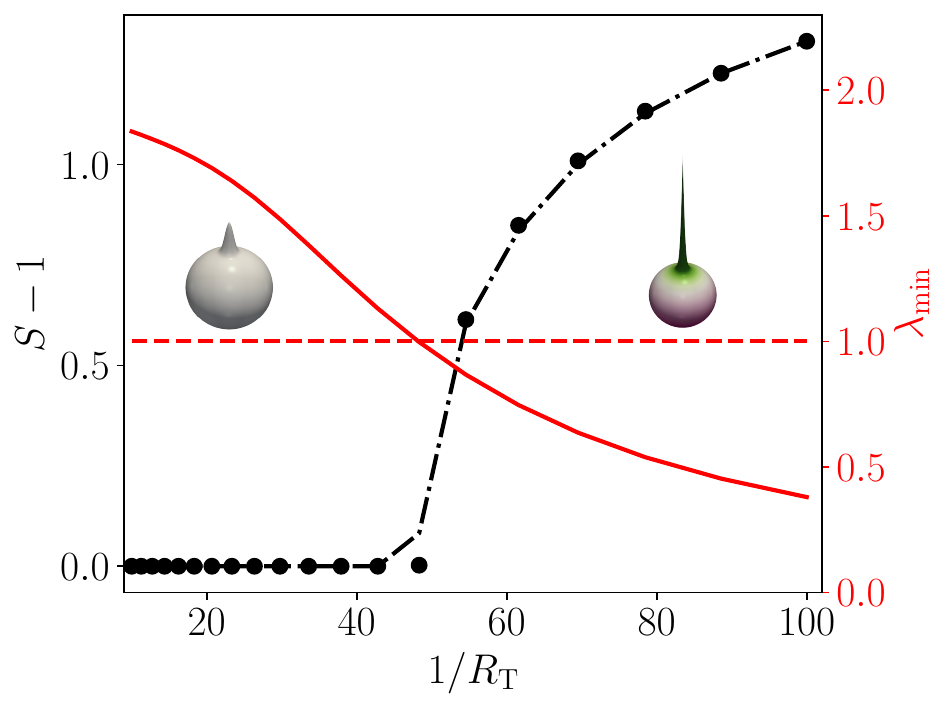}
\caption{Phase separation on tethers. The sorting ratio $S-1$ (points, left axis) as a function of the effective tether curvature $1/R_{T}$ (left axis), together with the weakly-nonlinear dependence $\sqrt{1-\lambda_{\text{min}}}$ (dashed line, fitted to the last point). We plot $\lambda_{\text{min}}$ in red (right axis) to demonstrate that the transition occurs once $\lambda_{\text{min}}\!\simeq\!1$ (dashed line). Insets show the mixed and phase separated tethers (left and right respectively).}
\label{fig:fig5}
\end{figure}

Finally, many experimental realizations use a micropipette, that allows pulling a nanotube from a spherical membrane (e.g., giant unilamellar vesicles)~\cite{sorre2009curvature,aimon2014membrane,roux2010membrane,sorre2012nature,tian2009sorting,beltran2019membrane}. The experimentally-reported sorting ratio $S$ represents a ratio of intensities between the fluorescence from the tube and the bulk --- here we define an equivalent sorting ratio that functions as a probe for the concentrations $S\!\equiv\!\frac{1 + |\phi_{\text{tube}}|}{1 - |\phi_{\text{bulk}}|}$~\cite{SM}. Some experiments report curvature-induced lipid phase separation~\cite{sorre2009curvature,beltran2019membrane} --- we numerically demonstrate in Fig.~\ref{fig:fig5} a similar effect. Instead of perturbing a sphere with a spherical harmonic, we deform it in a localized region, resulting in a tether-like geometry (see insets in Fig.~\ref{fig:fig5}). We plot the numerically-obtained $\lambda_{\text{min}}$ versus the effective curvature along the tether $1/R_{T}$ ($R_T$ being the tether radius). We also plot $S-1$ obtained from the numerical solutions on the tether geometries~\cite{SM}, and the predicted $\sqrt{1-\lambda_{\text{min}}}$ from the weakly-nonlinear analysis, clearly indicating that the geometric coupling via $\LB$ suffices to induce phase separations.

The resulting saturation profile provides a testable signature to differentiate geometric from mechanical sorting. Mechanical sorting models directly couple composition to the bending energy and can therefore produce curvature-dependent energetic preferences in addition to the geometric contribution identified here~\cite{aimon2014membrane}. The geometric model instead predicts that sorting continues monotonically with spectral softening. These distinct dependencies suggest that measurements at sufficiently high curvatures could help distinguish geometric and mechanical contributions.

\emph{Discussion. ---}
Biological cells and tissues adopt non-spherical shapes for prolonged periods during critical processes. While the canonical framework of phase separation in biological membranes couples the mixing free energy with the Helfrich energy for membrane bending, we demonstrated above that deformations lower the energetic cost of forming spatial variations (strengthening sorting). The softening of $\lambda_{\text{min}}$ under shape deformations destabilizes the homogeneous state (whereas the uniform distribution is globally stable in the spherical geometry).

Unlike standard Landau-Ginzburg approaches to phase transition ~\cite{kardar2007statistical,safran2018statistical}, the deformation-driven instability modifies the contribution of the gradient term through $\lambda_{\text{min}}$ and does not require a change in the local mixing free energy. On highly deformed spheres, the geometry softens the minimal eigenvalue, and phase separation proceeds via spontaneous symmetry breaking (Figs.~\ref{fig:fig3}-\ref{fig:fig4}), without an explicit mechanical coupling. The fact that the instability occurs near criticality, and that the gradient term is necessary align well with recent understanding of nucleation and growth on curved surfaces~\cite{gomez2015phase}, the experimental remarks that curvature-induced sorting occurred close to the critical concentration~\cite{sorre2009curvature}, and with the realization that interactions are needed to induce such separation~\cite{tian2009sorting}. 

Within the Landau-Ginzburg model, $\xi$ is the characteristic length associated with the competition between the local mixing and gradient terms, scaling as $\sqrt{\eta}$~\cite{safran2018statistical,SM}. For a sphere of radius $R$, $\lambda_{\text{min}}\!=\!\frac{2}{R^2}$, and the homogeneous state is stable when $R\!\lesssim\!\sqrt{2}\xi$ (for $\phi_0\!=\!0$). For a sufficiently large sphere $R\!\gg\!\xi$, phase separation already occurs on the undeformed sphere and deformation is not required. Geometry-induced separation is most relevant when the system size is comparable to the characteristic length, such that deformation can lower $\lambda_{\text{min}}$ across the instability threshold $\xi^2\lambda_{\text{min}}\!=\!1$.

The proposed deformation-induced instability provides an alternative framework for interpreting experimentally-observed phase separation on curved manifolds. However, because the purely geometric mechanism relies on spontaneous symmetry breaking, it does not inherently break the energetic degeneracy between equivalent spatial regions. The robust, deterministic localization frequently observed in biological systems implies the presence of additional energetic contributions. We suggest that even a weak explicit mechanical coupling or intrinsic curvature preference is required to act as a symmetry-breaking field. In this unified view, geometry controls the onset of demixing, while weak mechanical forces direct the resulting domains.

\emph{Acknowledgments. --- } This research was partially supported by NSF through the Princeton University (PCCM) Materials Research Science and Engineering Center DMR-2614599. A.M. acknowledges support from the \href{https://doi.org/10.37717/2021-3362}{James S. McDonnell Foundation Postdoctoral Fellowship Award in Complex Systems}. We thank A. Ko\v{s}mrlj, S. Mendez, and N. Wingreen for helpful comments and discussions.


\begin{thebibliography}{74}%
\makeatletter
\providecommand \@ifxundefined [1]{%
 \@ifx{#1\undefined}
}%
\providecommand \@ifnum [1]{%
 \ifnum #1\expandafter \@firstoftwo
 \else \expandafter \@secondoftwo
 \fi
}%
\providecommand \@ifx [1]{%
 \ifx #1\expandafter \@firstoftwo
 \else \expandafter \@secondoftwo
 \fi
}%
\providecommand \natexlab [1]{#1}%
\providecommand \enquote  [1]{``#1''}%
\providecommand \bibnamefont  [1]{#1}%
\providecommand \bibfnamefont [1]{#1}%
\providecommand \citenamefont [1]{#1}%
\providecommand \href@noop [0]{\@secondoftwo}%
\providecommand \href [0]{\begingroup \@sanitize@url \@href}%
\providecommand \@href[1]{\@@startlink{#1}\@@href}%
\providecommand \@@href[1]{\endgroup#1\@@endlink}%
\providecommand \@sanitize@url [0]{\catcode `\\12\catcode `\$12\catcode `\&12\catcode `\#12\catcode `\^12\catcode `\_12\catcode `\%12\relax}%
\providecommand \@@startlink[1]{}%
\providecommand \@@endlink[0]{}%
\providecommand \url  [0]{\begingroup\@sanitize@url \@url }%
\providecommand \@url [1]{\endgroup\@href {#1}{\urlprefix }}%
\providecommand \urlprefix  [0]{URL }%
\providecommand \Eprint [0]{\href }%
\providecommand \doibase [0]{https://doi.org/}%
\providecommand \selectlanguage [0]{\@gobble}%
\providecommand \bibinfo  [0]{\@secondoftwo}%
\providecommand \bibfield  [0]{\@secondoftwo}%
\providecommand \translation [1]{[#1]}%
\providecommand \BibitemOpen [0]{}%
\providecommand \bibitemStop [0]{}%
\providecommand \bibitemNoStop [0]{.\EOS\space}%
\providecommand \EOS [0]{\spacefactor3000\relax}%
\providecommand \BibitemShut  [1]{\csname bibitem#1\endcsname}%
\let\auto@bib@innerbib\@empty
\bibitem [{\citenamefont {Lindblom}\ and\ \citenamefont {Or{\"a}dd}(2009)}]{lindblom2009lipid}%
  \BibitemOpen
  \bibfield  {author} {\bibinfo {author} {\bibfnamefont {G.}~\bibnamefont {Lindblom}}\ and\ \bibinfo {author} {\bibfnamefont {G.}~\bibnamefont {Or{\"a}dd}},\ }\href {https://doi.org/10.1016/j.bbamem.2008.08.016} {\bibfield  {journal} {\bibinfo  {journal} {Biochim. Biophys. Acta, Biomembr.}\ }\textbf {\bibinfo {volume} {1788}},\ \bibinfo {pages} {234} (\bibinfo {year} {2009})}\BibitemShut {NoStop}%
\bibitem [{\citenamefont {Almeida}\ and\ \citenamefont {Vaz}(1995)}]{ALMEIDA1995305}%
  \BibitemOpen
  \bibfield  {author} {\bibinfo {author} {\bibfnamefont {P.}~\bibnamefont {Almeida}}\ and\ \bibinfo {author} {\bibfnamefont {W.}~\bibnamefont {Vaz}},\ }in\ \href {https://doi.org/https://doi.org/10.1016/S1383-8121(06)80023-0} {\emph {\bibinfo {booktitle} {Structure and Dynamics of Membranes}}},\ \bibinfo {series} {Handbook of Biological Physics}, Vol.~\bibinfo {volume} {1},\ \bibinfo {editor} {edited by\ \bibinfo {editor} {\bibfnamefont {R.}~\bibnamefont {Lipowsky}}\ and\ \bibinfo {editor} {\bibfnamefont {E.}~\bibnamefont {Sackmann}}}\ (\bibinfo  {publisher} {North-Holland},\ \bibinfo {year} {1995})\ pp.\ \bibinfo {pages} {305--357}\BibitemShut {NoStop}%
\bibitem [{\citenamefont {Kusumi}\ \emph {et~al.}(2012)\citenamefont {Kusumi}, \citenamefont {Fujiwara}, \citenamefont {Chadda}, \citenamefont {Xie}, \citenamefont {Tsunoyama}, \citenamefont {Kalay}, \citenamefont {Kasai},\ and\ \citenamefont {Suzuki}}]{kusumi2012dynamic}%
  \BibitemOpen
  \bibfield  {author} {\bibinfo {author} {\bibfnamefont {A.}~\bibnamefont {Kusumi}}, \bibinfo {author} {\bibfnamefont {T.~K.}\ \bibnamefont {Fujiwara}}, \bibinfo {author} {\bibfnamefont {R.}~\bibnamefont {Chadda}}, \bibinfo {author} {\bibfnamefont {M.}~\bibnamefont {Xie}}, \bibinfo {author} {\bibfnamefont {T.~A.}\ \bibnamefont {Tsunoyama}}, \bibinfo {author} {\bibfnamefont {Z.}~\bibnamefont {Kalay}}, \bibinfo {author} {\bibfnamefont {R.~S.}\ \bibnamefont {Kasai}},\ and\ \bibinfo {author} {\bibfnamefont {K.~G.~N.}\ \bibnamefont {Suzuki}},\ }\href {https://doi.org/10.1146/annurev-cellbio-100809-151736} {\bibfield  {journal} {\bibinfo  {journal} {Annu. Rev. Cell Dev. Biol.}\ }\textbf {\bibinfo {volume} {28}},\ \bibinfo {pages} {215} (\bibinfo {year} {2012})}\BibitemShut {NoStop}%
\bibitem [{\citenamefont {Axelrod}(1983)}]{axelrod1983lateral}%
  \BibitemOpen
  \bibfield  {author} {\bibinfo {author} {\bibfnamefont {D.}~\bibnamefont {Axelrod}},\ }\href {https://doi.org/10.1007/BF01870794} {\bibfield  {journal} {\bibinfo  {journal} {J. Membr. Biol.}\ }\textbf {\bibinfo {volume} {75}},\ \bibinfo {pages} {1} (\bibinfo {year} {1983})}\BibitemShut {NoStop}%
\bibitem [{\citenamefont {Saxton}(1999)}]{SAXTON1999229}%
  \BibitemOpen
  \bibfield  {author} {\bibinfo {author} {\bibfnamefont {M.~J.}\ \bibnamefont {Saxton}},\ }in\ \href {https://doi.org/https://doi.org/10.1016/S0070-2161(08)61046-1} {\emph {\bibinfo {booktitle} {Membrane Permeability}}},\ \bibinfo {series} {Current Topics in Membranes}, Vol.~\bibinfo {volume} {48},\ \bibinfo {editor} {edited by\ \bibinfo {editor} {\bibfnamefont {D.~W.}\ \bibnamefont {Deamer}}, \bibinfo {editor} {\bibfnamefont {A.}~\bibnamefont {Kleinzeller}},\ and\ \bibinfo {editor} {\bibfnamefont {D.~M.}\ \bibnamefont {Fambrough}}}\ (\bibinfo  {publisher} {Academic Press},\ \bibinfo {year} {1999})\ pp.\ \bibinfo {pages} {229--282}\BibitemShut {NoStop}%
\bibitem [{\citenamefont {Singer}(1971)}]{singer1971molecular}%
  \BibitemOpen
  \bibfield  {author} {\bibinfo {author} {\bibfnamefont {S.~J.}\ \bibnamefont {Singer}},\ }in\ \href {https://doi.org/10.1016/B978-0-12-598650-2.50009-0} {\emph {\bibinfo {booktitle} {Structure and function of biological membranes}}}\ (\bibinfo  {publisher} {Elsevier},\ \bibinfo {year} {1971})\ pp.\ \bibinfo {pages} {145--222}\BibitemShut {NoStop}%
\bibitem [{\citenamefont {Singer}\ and\ \citenamefont {Nicolson}(1972)}]{singer1972fluid}%
  \BibitemOpen
  \bibfield  {author} {\bibinfo {author} {\bibfnamefont {S.~J.}\ \bibnamefont {Singer}}\ and\ \bibinfo {author} {\bibfnamefont {G.~L.}\ \bibnamefont {Nicolson}},\ }\href {https://doi.org/10.1126/science.175.4023.720} {\bibfield  {journal} {\bibinfo  {journal} {Science}\ }\textbf {\bibinfo {volume} {175}},\ \bibinfo {pages} {720} (\bibinfo {year} {1972})}\BibitemShut {NoStop}%
\bibitem [{\citenamefont {Singer}(1974)}]{singer1974molecular}%
  \BibitemOpen
  \bibfield  {author} {\bibinfo {author} {\bibfnamefont {S.~J.}\ \bibnamefont {Singer}},\ }\href {https://doi.org/10.1146/annurev.bi.43.070174.004105} {\bibfield  {journal} {\bibinfo  {journal} {Annu. Rev. Biochem.}\ }\textbf {\bibinfo {volume} {43}},\ \bibinfo {pages} {805} (\bibinfo {year} {1974})}\BibitemShut {NoStop}%
\bibitem [{\citenamefont {Lingwood}\ and\ \citenamefont {Simons}(2010)}]{lingwood2010lipid}%
  \BibitemOpen
  \bibfield  {author} {\bibinfo {author} {\bibfnamefont {D.}~\bibnamefont {Lingwood}}\ and\ \bibinfo {author} {\bibfnamefont {K.}~\bibnamefont {Simons}},\ }\href {https://doi.org/10.1126/science.1174621} {\bibfield  {journal} {\bibinfo  {journal} {Science}\ }\textbf {\bibinfo {volume} {327}},\ \bibinfo {pages} {46} (\bibinfo {year} {2010})}\BibitemShut {NoStop}%
\bibitem [{\citenamefont {Engelman}(2005)}]{engelman2005membranes}%
  \BibitemOpen
  \bibfield  {author} {\bibinfo {author} {\bibfnamefont {D.~M.}\ \bibnamefont {Engelman}},\ }\href {https://doi.org/10.1038/nature04394} {\bibfield  {journal} {\bibinfo  {journal} {Nature}\ }\textbf {\bibinfo {volume} {438}} (\bibinfo {year} {2005})}\BibitemShut {NoStop}%
\bibitem [{\citenamefont {Peters}\ and\ \citenamefont {Cherry}(1982)}]{peters1982lateral}%
  \BibitemOpen
  \bibfield  {author} {\bibinfo {author} {\bibfnamefont {R.}~\bibnamefont {Peters}}\ and\ \bibinfo {author} {\bibfnamefont {R.~J.}\ \bibnamefont {Cherry}},\ }\href {https://doi.org/10.1073/pnas.79.14.4317} {\bibfield  {journal} {\bibinfo  {journal} {Proc. Natl. Acad. Sci. U.S.A.}\ }\textbf {\bibinfo {volume} {79}},\ \bibinfo {pages} {4317} (\bibinfo {year} {1982})}\BibitemShut {NoStop}%
\bibitem [{\citenamefont {Gambin}\ \emph {et~al.}(2006)\citenamefont {Gambin}, \citenamefont {Lopez-Esparza}, \citenamefont {Reffay}, \citenamefont {Sierecki}, \citenamefont {Gov}, \citenamefont {Genest}, \citenamefont {Hodges},\ and\ \citenamefont {Urbach}}]{gambin2006lateral}%
  \BibitemOpen
  \bibfield  {author} {\bibinfo {author} {\bibfnamefont {Y.}~\bibnamefont {Gambin}}, \bibinfo {author} {\bibfnamefont {R.}~\bibnamefont {Lopez-Esparza}}, \bibinfo {author} {\bibfnamefont {M.}~\bibnamefont {Reffay}}, \bibinfo {author} {\bibfnamefont {E.}~\bibnamefont {Sierecki}}, \bibinfo {author} {\bibfnamefont {N.~S.}\ \bibnamefont {Gov}}, \bibinfo {author} {\bibfnamefont {M.~H. R.~S.}\ \bibnamefont {Genest}}, \bibinfo {author} {\bibfnamefont {R.~S.}\ \bibnamefont {Hodges}},\ and\ \bibinfo {author} {\bibfnamefont {W.}~\bibnamefont {Urbach}},\ }\href {https://doi.org/10.1073/pnas.0511026103} {\bibfield  {journal} {\bibinfo  {journal} {Proc. Natl. Acad. Sci. U.S.A.}\ }\textbf {\bibinfo {volume} {103}},\ \bibinfo {pages} {2098} (\bibinfo {year} {2006})}\BibitemShut {NoStop}%
\bibitem [{\citenamefont {Saffman}\ and\ \citenamefont {Delbr{\"u}ck}(1975)}]{saffman1975brownian}%
  \BibitemOpen
  \bibfield  {author} {\bibinfo {author} {\bibfnamefont {P.~G.}\ \bibnamefont {Saffman}}\ and\ \bibinfo {author} {\bibfnamefont {M.}~\bibnamefont {Delbr{\"u}ck}},\ }\href {https://doi.org/10.1073/pnas.72.8.3111} {\bibfield  {journal} {\bibinfo  {journal} {Proc. Natl. Acad. Sci. U.S.A.}\ }\textbf {\bibinfo {volume} {72}},\ \bibinfo {pages} {3111} (\bibinfo {year} {1975})}\BibitemShut {NoStop}%
\bibitem [{\citenamefont {Henle}\ and\ \citenamefont {Levine}(2010)}]{henle2010hydrodynamics}%
  \BibitemOpen
  \bibfield  {author} {\bibinfo {author} {\bibfnamefont {M.~L.}\ \bibnamefont {Henle}}\ and\ \bibinfo {author} {\bibfnamefont {A.~J.}\ \bibnamefont {Levine}},\ }\href {https://doi.org/10.1103/PhysRevE.81.011905} {\bibfield  {journal} {\bibinfo  {journal} {Phys. Rev. E}\ }\textbf {\bibinfo {volume} {81}},\ \bibinfo {pages} {011905} (\bibinfo {year} {2010})}\BibitemShut {NoStop}%
\bibitem [{\citenamefont {Phillips}\ \emph {et~al.}(2009)\citenamefont {Phillips}, \citenamefont {Ursell}, \citenamefont {Wiggins},\ and\ \citenamefont {Sens}}]{phillips2009emerging}%
  \BibitemOpen
  \bibfield  {author} {\bibinfo {author} {\bibfnamefont {R.}~\bibnamefont {Phillips}}, \bibinfo {author} {\bibfnamefont {T.}~\bibnamefont {Ursell}}, \bibinfo {author} {\bibfnamefont {P.}~\bibnamefont {Wiggins}},\ and\ \bibinfo {author} {\bibfnamefont {P.}~\bibnamefont {Sens}},\ }\href {https://doi.org/10.1038/nature08147} {\bibfield  {journal} {\bibinfo  {journal} {Nature}\ }\textbf {\bibinfo {volume} {459}},\ \bibinfo {pages} {379} (\bibinfo {year} {2009})}\BibitemShut {NoStop}%
\bibitem [{\citenamefont {Baumgart}\ \emph {et~al.}(2003)\citenamefont {Baumgart}, \citenamefont {Hess},\ and\ \citenamefont {Webb}}]{baumgart2003imaging}%
  \BibitemOpen
  \bibfield  {author} {\bibinfo {author} {\bibfnamefont {T.}~\bibnamefont {Baumgart}}, \bibinfo {author} {\bibfnamefont {S.~T.}\ \bibnamefont {Hess}},\ and\ \bibinfo {author} {\bibfnamefont {W.~W.}\ \bibnamefont {Webb}},\ }\href {https://doi.org/10.1038/nature02013} {\bibfield  {journal} {\bibinfo  {journal} {Nature}\ }\textbf {\bibinfo {volume} {425}},\ \bibinfo {pages} {821} (\bibinfo {year} {2003})}\BibitemShut {NoStop}%
\bibitem [{\citenamefont {Bhatia}\ \emph {et~al.}(2009)\citenamefont {Bhatia}, \citenamefont {Madsen}, \citenamefont {Bolinger}, \citenamefont {Kunding}, \citenamefont {Hedeg{\aa}rd}, \citenamefont {Gether},\ and\ \citenamefont {Stamou}}]{bhatia2009amphipathic}%
  \BibitemOpen
  \bibfield  {author} {\bibinfo {author} {\bibfnamefont {V.~K.}\ \bibnamefont {Bhatia}}, \bibinfo {author} {\bibfnamefont {K.~L.}\ \bibnamefont {Madsen}}, \bibinfo {author} {\bibfnamefont {P.~Y.}\ \bibnamefont {Bolinger}}, \bibinfo {author} {\bibfnamefont {A.}~\bibnamefont {Kunding}}, \bibinfo {author} {\bibfnamefont {P.}~\bibnamefont {Hedeg{\aa}rd}}, \bibinfo {author} {\bibfnamefont {U.}~\bibnamefont {Gether}},\ and\ \bibinfo {author} {\bibfnamefont {D.}~\bibnamefont {Stamou}},\ }\href {https://link.springer.com/article/10.1038/emboj.2009.261} {\bibfield  {journal} {\bibinfo  {journal} {EMBO J.}\ }\textbf {\bibinfo {volume} {28}},\ \bibinfo {pages} {3303} (\bibinfo {year} {2009})}\BibitemShut {NoStop}%
\bibitem [{\citenamefont {Hatzakis}\ \emph {et~al.}(2009)\citenamefont {Hatzakis}, \citenamefont {Bhatia}, \citenamefont {Larsen}, \citenamefont {Madsen}, \citenamefont {Bolinger}, \citenamefont {Kunding}, \citenamefont {Castillo}, \citenamefont {Gether}, \citenamefont {Hedeg{\aa}rd},\ and\ \citenamefont {Stamou}}]{hatzakis2009curved}%
  \BibitemOpen
  \bibfield  {author} {\bibinfo {author} {\bibfnamefont {N.~S.}\ \bibnamefont {Hatzakis}}, \bibinfo {author} {\bibfnamefont {V.~K.}\ \bibnamefont {Bhatia}}, \bibinfo {author} {\bibfnamefont {J.}~\bibnamefont {Larsen}}, \bibinfo {author} {\bibfnamefont {K.~L.}\ \bibnamefont {Madsen}}, \bibinfo {author} {\bibfnamefont {P.~Y.}\ \bibnamefont {Bolinger}}, \bibinfo {author} {\bibfnamefont {A.~H.}\ \bibnamefont {Kunding}}, \bibinfo {author} {\bibfnamefont {J.}~\bibnamefont {Castillo}}, \bibinfo {author} {\bibfnamefont {U.}~\bibnamefont {Gether}}, \bibinfo {author} {\bibfnamefont {P.}~\bibnamefont {Hedeg{\aa}rd}},\ and\ \bibinfo {author} {\bibfnamefont {D.}~\bibnamefont {Stamou}},\ }\href {https://doi.org/10.1038/nchembio.213} {\bibfield  {journal} {\bibinfo  {journal} {Nat. Chem. Biol.}\ }\textbf {\bibinfo {volume} {5}},\ \bibinfo {pages} {835} (\bibinfo {year} {2009})}\BibitemShut {NoStop}%
\bibitem [{\citenamefont {Parthasarathy}\ and\ \citenamefont {Groves}(2006)}]{parthasarathy2006curvaturereview}%
  \BibitemOpen
  \bibfield  {author} {\bibinfo {author} {\bibfnamefont {R.}~\bibnamefont {Parthasarathy}}\ and\ \bibinfo {author} {\bibfnamefont {J.~T.}\ \bibnamefont {Groves}},\ }\href {https://doi.org/10.1039/b608631d} {\bibfield  {journal} {\bibinfo  {journal} {Soft Matter}\ }\textbf {\bibinfo {volume} {3}},\ \bibinfo {pages} {24} (\bibinfo {year} {2006})}\BibitemShut {NoStop}%
\bibitem [{\citenamefont {Heinrich}\ \emph {et~al.}(2010)\citenamefont {Heinrich}, \citenamefont {Tian}, \citenamefont {Esposito},\ and\ \citenamefont {Baumgart}}]{heinrich2010dynamic}%
  \BibitemOpen
  \bibfield  {author} {\bibinfo {author} {\bibfnamefont {M.}~\bibnamefont {Heinrich}}, \bibinfo {author} {\bibfnamefont {A.}~\bibnamefont {Tian}}, \bibinfo {author} {\bibfnamefont {C.}~\bibnamefont {Esposito}},\ and\ \bibinfo {author} {\bibfnamefont {T.}~\bibnamefont {Baumgart}},\ }\href {https://doi.org/10.1073/pnas.0913997107} {\bibfield  {journal} {\bibinfo  {journal} {Proc. Natl. Acad. Sci. U.S.A.}\ }\textbf {\bibinfo {volume} {107}},\ \bibinfo {pages} {7208} (\bibinfo {year} {2010})}\BibitemShut {NoStop}%
\bibitem [{\citenamefont {Zimmerberg}\ and\ \citenamefont {Kozlov}(2006)}]{Zimmerberg2006}%
  \BibitemOpen
  \bibfield  {author} {\bibinfo {author} {\bibfnamefont {J.}~\bibnamefont {Zimmerberg}}\ and\ \bibinfo {author} {\bibfnamefont {M.~M.}\ \bibnamefont {Kozlov}},\ }\href {https://doi.org/10.1038/nrm1784} {\bibfield  {journal} {\bibinfo  {journal} {Nat. Rev. Mol. Cell Biol.}\ }\textbf {\bibinfo {volume} {7}},\ \bibinfo {pages} {9} (\bibinfo {year} {2006})}\BibitemShut {NoStop}%
\bibitem [{\citenamefont {Kawakatsu}\ \emph {et~al.}(1993)\citenamefont {Kawakatsu}, \citenamefont {Andelman}, \citenamefont {Kawasaki},\ and\ \citenamefont {Taniguchi}}]{kawakatsu1993phase}%
  \BibitemOpen
  \bibfield  {author} {\bibinfo {author} {\bibfnamefont {T.}~\bibnamefont {Kawakatsu}}, \bibinfo {author} {\bibfnamefont {D.}~\bibnamefont {Andelman}}, \bibinfo {author} {\bibfnamefont {K.}~\bibnamefont {Kawasaki}},\ and\ \bibinfo {author} {\bibfnamefont {T.}~\bibnamefont {Taniguchi}},\ }\href {https://doi.org/10.1051/jp2:1993177} {\bibfield  {journal} {\bibinfo  {journal} {J. Phys. II}\ }\textbf {\bibinfo {volume} {3}},\ \bibinfo {pages} {971} (\bibinfo {year} {1993})}\BibitemShut {NoStop}%
\bibitem [{\citenamefont {Seifert}\ and\ \citenamefont {Langer}(1993)}]{seifert1993viscous}%
  \BibitemOpen
  \bibfield  {author} {\bibinfo {author} {\bibfnamefont {U.}~\bibnamefont {Seifert}}\ and\ \bibinfo {author} {\bibfnamefont {S.~A.}\ \bibnamefont {Langer}},\ }\href {https://doi.org/10.1209/0295-5075/23/1/012} {\bibfield  {journal} {\bibinfo  {journal} {Europhys. Lett.}\ }\textbf {\bibinfo {volume} {23}},\ \bibinfo {pages} {71} (\bibinfo {year} {1993})}\BibitemShut {NoStop}%
\bibitem [{\citenamefont {Bassereau}\ \emph {et~al.}(2018)\citenamefont {Bassereau}, \citenamefont {Jin}, \citenamefont {Baumgart}, \citenamefont {Deserno}, \citenamefont {Dimova}, \citenamefont {Frolov}, \citenamefont {Bashkirov}, \citenamefont {Grubm{\"u}ller}, \citenamefont {Jahn}, \citenamefont {Risselada} \emph {et~al.}}]{bassereau2018}%
  \BibitemOpen
  \bibfield  {author} {\bibinfo {author} {\bibfnamefont {P.}~\bibnamefont {Bassereau}}, \bibinfo {author} {\bibfnamefont {R.}~\bibnamefont {Jin}}, \bibinfo {author} {\bibfnamefont {T.}~\bibnamefont {Baumgart}}, \bibinfo {author} {\bibfnamefont {M.}~\bibnamefont {Deserno}}, \bibinfo {author} {\bibfnamefont {R.}~\bibnamefont {Dimova}}, \bibinfo {author} {\bibfnamefont {V.~A.}\ \bibnamefont {Frolov}}, \bibinfo {author} {\bibfnamefont {P.~V.}\ \bibnamefont {Bashkirov}}, \bibinfo {author} {\bibfnamefont {H.}~\bibnamefont {Grubm{\"u}ller}}, \bibinfo {author} {\bibfnamefont {R.}~\bibnamefont {Jahn}}, \bibinfo {author} {\bibfnamefont {H.~J.}\ \bibnamefont {Risselada}}, \emph {et~al.},\ }\href {https://doi.org/10.1088/1361-6463/aacb98} {\bibfield  {journal} {\bibinfo  {journal} {J. Phys. D: Appl. Phys.}\ }\textbf {\bibinfo {volume} {51}},\ \bibinfo {pages} {343001} (\bibinfo {year} {2018})}\BibitemShut {NoStop}%
\bibitem [{\citenamefont {Netz}\ and\ \citenamefont {Pincus}(1995)}]{netz1995inhomogeneous}%
  \BibitemOpen
  \bibfield  {author} {\bibinfo {author} {\bibfnamefont {R.~R.}\ \bibnamefont {Netz}}\ and\ \bibinfo {author} {\bibfnamefont {P.}~\bibnamefont {Pincus}},\ }\href {https://doi.org/10.1103/PhysRevE.52.4114} {\bibfield  {journal} {\bibinfo  {journal} {Phys. Rev. E}\ }\textbf {\bibinfo {volume} {52}},\ \bibinfo {pages} {4114} (\bibinfo {year} {1995})}\BibitemShut {NoStop}%
\bibitem [{\citenamefont {Helfrich}(1973)}]{helfrich1973elastic}%
  \BibitemOpen
  \bibfield  {author} {\bibinfo {author} {\bibfnamefont {W.}~\bibnamefont {Helfrich}},\ }\href {https://doi.org/10.1515/znc-1973-11-1209} {\bibfield  {journal} {\bibinfo  {journal} {Z. Naturforsch.}\ }\textbf {\bibinfo {volume} {28}},\ \bibinfo {pages} {693} (\bibinfo {year} {1973})}\BibitemShut {NoStop}%
\bibitem [{\citenamefont {Winter}\ \emph {et~al.}(2025)\citenamefont {Winter}, \citenamefont {Liu}, \citenamefont {Ziepke}, \citenamefont {Dadunashvili},\ and\ \citenamefont {Frey}}]{winter2025phase}%
  \BibitemOpen
  \bibfield  {author} {\bibinfo {author} {\bibfnamefont {A.}~\bibnamefont {Winter}}, \bibinfo {author} {\bibfnamefont {Y.}~\bibnamefont {Liu}}, \bibinfo {author} {\bibfnamefont {A.}~\bibnamefont {Ziepke}}, \bibinfo {author} {\bibfnamefont {G.}~\bibnamefont {Dadunashvili}},\ and\ \bibinfo {author} {\bibfnamefont {E.}~\bibnamefont {Frey}},\ }\href {https://doi.org/10.1103/PhysRevE.111.044405} {\bibfield  {journal} {\bibinfo  {journal} {Phys. Rev. E}\ }\textbf {\bibinfo {volume} {111}},\ \bibinfo {pages} {044405} (\bibinfo {year} {2025})}\BibitemShut {NoStop}%
\bibitem [{\citenamefont {J{\"u}licher}\ and\ \citenamefont {Lipowsky}(1996)}]{julicher1996shape}%
  \BibitemOpen
  \bibfield  {author} {\bibinfo {author} {\bibfnamefont {F.}~\bibnamefont {J{\"u}licher}}\ and\ \bibinfo {author} {\bibfnamefont {R.}~\bibnamefont {Lipowsky}},\ }\href {https://doi.org/10.1103/PhysRevE.53.2670} {\bibfield  {journal} {\bibinfo  {journal} {Phys. Rev. E}\ }\textbf {\bibinfo {volume} {53}},\ \bibinfo {pages} {2670} (\bibinfo {year} {1996})}\BibitemShut {NoStop}%
\bibitem [{\citenamefont {J{\"u}licher}\ and\ \citenamefont {Lipowsky}(1993)}]{julicher1993domain}%
  \BibitemOpen
  \bibfield  {author} {\bibinfo {author} {\bibfnamefont {F.}~\bibnamefont {J{\"u}licher}}\ and\ \bibinfo {author} {\bibfnamefont {R.}~\bibnamefont {Lipowsky}},\ }\href {https://doi.org/10.1103/PhysRevLett.70.2964} {\bibfield  {journal} {\bibinfo  {journal} {Phys. Rev. Lett.}\ }\textbf {\bibinfo {volume} {70}},\ \bibinfo {pages} {2964} (\bibinfo {year} {1993})}\BibitemShut {NoStop}%
\bibitem [{\citenamefont {Leibler}(1986)}]{leibler1986curvature}%
  \BibitemOpen
  \bibfield  {author} {\bibinfo {author} {\bibfnamefont {S.}~\bibnamefont {Leibler}},\ }\href {https://doi.org/10.1051/jphys:01986004703050700} {\bibfield  {journal} {\bibinfo  {journal} {J. Phys. (Paris)}\ }\textbf {\bibinfo {volume} {47}},\ \bibinfo {pages} {507} (\bibinfo {year} {1986})}\BibitemShut {NoStop}%
\bibitem [{\citenamefont {Seifert}(1993)}]{seifert1993curvature}%
  \BibitemOpen
  \bibfield  {author} {\bibinfo {author} {\bibfnamefont {U.}~\bibnamefont {Seifert}},\ }\href {https://doi.org/10.1103/PhysRevLett.70.1335} {\bibfield  {journal} {\bibinfo  {journal} {Phys. Rev. Lett.}\ }\textbf {\bibinfo {volume} {70}},\ \bibinfo {pages} {1335} (\bibinfo {year} {1993})}\BibitemShut {NoStop}%
\bibitem [{\citenamefont {Seifert}(1997)}]{seifert1997configurations}%
  \BibitemOpen
  \bibfield  {author} {\bibinfo {author} {\bibfnamefont {U.}~\bibnamefont {Seifert}},\ }\href {https://doi.org/10.1080/00018739700101488} {\bibfield  {journal} {\bibinfo  {journal} {Adv. Phys.}\ }\textbf {\bibinfo {volume} {46}},\ \bibinfo {pages} {13} (\bibinfo {year} {1997})}\BibitemShut {NoStop}%
\bibitem [{\citenamefont {Reister-Gottfried}\ \emph {et~al.}(2010)\citenamefont {Reister-Gottfried}, \citenamefont {Leitenberger},\ and\ \citenamefont {Seifert}}]{reister2010diffusing}%
  \BibitemOpen
  \bibfield  {author} {\bibinfo {author} {\bibfnamefont {E.}~\bibnamefont {Reister-Gottfried}}, \bibinfo {author} {\bibfnamefont {S.~M.}\ \bibnamefont {Leitenberger}},\ and\ \bibinfo {author} {\bibfnamefont {U.}~\bibnamefont {Seifert}},\ }\href {https://doi.org/10.1103/PhysRevE.81.031903} {\bibfield  {journal} {\bibinfo  {journal} {Phys. Rev. E}\ }\textbf {\bibinfo {volume} {81}},\ \bibinfo {pages} {031903} (\bibinfo {year} {2010})}\BibitemShut {NoStop}%
\bibitem [{\citenamefont {Huang}\ \emph {et~al.}(2006)\citenamefont {Huang}, \citenamefont {Mukhopadhyay},\ and\ \citenamefont {Wingreen}}]{huang2006curvature}%
  \BibitemOpen
  \bibfield  {author} {\bibinfo {author} {\bibfnamefont {K.~C.}\ \bibnamefont {Huang}}, \bibinfo {author} {\bibfnamefont {R.}~\bibnamefont {Mukhopadhyay}},\ and\ \bibinfo {author} {\bibfnamefont {N.~S.}\ \bibnamefont {Wingreen}},\ }\href {https://doi.org/10.1371/journal.pcbi.0020151} {\bibfield  {journal} {\bibinfo  {journal} {PLoS Comput. Biol.}\ }\textbf {\bibinfo {volume} {2}},\ \bibinfo {pages} {e151} (\bibinfo {year} {2006})}\BibitemShut {NoStop}%
\bibitem [{\citenamefont {Mukhopadhyay}\ \emph {et~al.}(2008)\citenamefont {Mukhopadhyay}, \citenamefont {Huang},\ and\ \citenamefont {Wingreen}}]{mukhopadhyay2008lipid}%
  \BibitemOpen
  \bibfield  {author} {\bibinfo {author} {\bibfnamefont {R.}~\bibnamefont {Mukhopadhyay}}, \bibinfo {author} {\bibfnamefont {K.~C.}\ \bibnamefont {Huang}},\ and\ \bibinfo {author} {\bibfnamefont {N.~S.}\ \bibnamefont {Wingreen}},\ }\href {https://doi.org/10.1529/biophysj.107.126920} {\bibfield  {journal} {\bibinfo  {journal} {Biophys. J.}\ }\textbf {\bibinfo {volume} {95}},\ \bibinfo {pages} {1034} (\bibinfo {year} {2008})}\BibitemShut {NoStop}%
\bibitem [{\citenamefont {Jiang}\ and\ \citenamefont {Powers}(2008)}]{jiang2008curvature}%
  \BibitemOpen
  \bibfield  {author} {\bibinfo {author} {\bibfnamefont {H.}~\bibnamefont {Jiang}}\ and\ \bibinfo {author} {\bibfnamefont {T.~R.}\ \bibnamefont {Powers}},\ }\href {https://doi.org/10.1103/PhysRevLett.101.018103} {\bibfield  {journal} {\bibinfo  {journal} {Phys. Rev. Lett.}\ }\textbf {\bibinfo {volume} {101}},\ \bibinfo {pages} {018103} (\bibinfo {year} {2008})}\BibitemShut {NoStop}%
\bibitem [{\citenamefont {Jiang}(2012)}]{jiang2012dynamic}%
  \BibitemOpen
  \bibfield  {author} {\bibinfo {author} {\bibfnamefont {H.}~\bibnamefont {Jiang}},\ }\href {https://doi.org/10.1103/PhysRevLett.109.198101} {\bibfield  {journal} {\bibinfo  {journal} {Phys. Rev. Lett.}\ }\textbf {\bibinfo {volume} {109}},\ \bibinfo {pages} {198101} (\bibinfo {year} {2012})}\BibitemShut {NoStop}%
\bibitem [{\citenamefont {Reynwar}\ \emph {et~al.}(2007)\citenamefont {Reynwar}, \citenamefont {Illya}, \citenamefont {Harmandaris}, \citenamefont {M{\"u}ller}, \citenamefont {Kremer},\ and\ \citenamefont {Deserno}}]{reynwar2007aggregation}%
  \BibitemOpen
  \bibfield  {author} {\bibinfo {author} {\bibfnamefont {B.~J.}\ \bibnamefont {Reynwar}}, \bibinfo {author} {\bibfnamefont {G.}~\bibnamefont {Illya}}, \bibinfo {author} {\bibfnamefont {V.~A.}\ \bibnamefont {Harmandaris}}, \bibinfo {author} {\bibfnamefont {M.~M.}\ \bibnamefont {M{\"u}ller}}, \bibinfo {author} {\bibfnamefont {K.}~\bibnamefont {Kremer}},\ and\ \bibinfo {author} {\bibfnamefont {M.}~\bibnamefont {Deserno}},\ }\href {https://doi.org/10.1038/nature05840} {\bibfield  {journal} {\bibinfo  {journal} {Nature}\ }\textbf {\bibinfo {volume} {447}},\ \bibinfo {pages} {461} (\bibinfo {year} {2007})}\BibitemShut {NoStop}%
\bibitem [{\citenamefont {Aimon}\ \emph {et~al.}(2014)\citenamefont {Aimon}, \citenamefont {Callan-Jones}, \citenamefont {Berthaud}, \citenamefont {Pinot}, \citenamefont {Toombes},\ and\ \citenamefont {Bassereau}}]{aimon2014membrane}%
  \BibitemOpen
  \bibfield  {author} {\bibinfo {author} {\bibfnamefont {S.}~\bibnamefont {Aimon}}, \bibinfo {author} {\bibfnamefont {A.}~\bibnamefont {Callan-Jones}}, \bibinfo {author} {\bibfnamefont {A.}~\bibnamefont {Berthaud}}, \bibinfo {author} {\bibfnamefont {M.}~\bibnamefont {Pinot}}, \bibinfo {author} {\bibfnamefont {G.~E.~S.}\ \bibnamefont {Toombes}},\ and\ \bibinfo {author} {\bibfnamefont {P.}~\bibnamefont {Bassereau}},\ }\href {https://doi.org/10.1016/j.devcel.2013.12.01} {\bibfield  {journal} {\bibinfo  {journal} {Dev. Cell}\ }\textbf {\bibinfo {volume} {28}},\ \bibinfo {pages} {212} (\bibinfo {year} {2014})}\BibitemShut {NoStop}%
\bibitem [{\citenamefont {Roux}\ \emph {et~al.}(2010)\citenamefont {Roux}, \citenamefont {Koster}, \citenamefont {Lenz}, \citenamefont {Sorre}, \citenamefont {Manneville}, \citenamefont {Nassoy},\ and\ \citenamefont {Bassereau}}]{roux2010membrane}%
  \BibitemOpen
  \bibfield  {author} {\bibinfo {author} {\bibfnamefont {A.}~\bibnamefont {Roux}}, \bibinfo {author} {\bibfnamefont {G.}~\bibnamefont {Koster}}, \bibinfo {author} {\bibfnamefont {M.}~\bibnamefont {Lenz}}, \bibinfo {author} {\bibfnamefont {B.}~\bibnamefont {Sorre}}, \bibinfo {author} {\bibfnamefont {J.~B.}\ \bibnamefont {Manneville}}, \bibinfo {author} {\bibfnamefont {P.}~\bibnamefont {Nassoy}},\ and\ \bibinfo {author} {\bibfnamefont {P.}~\bibnamefont {Bassereau}},\ }\href {https://doi.org/10.1073/pnas.0913734107} {\bibfield  {journal} {\bibinfo  {journal} {Proc. Natl. Acad. Sci. U.S.A.}\ }\textbf {\bibinfo {volume} {107}},\ \bibinfo {pages} {4141} (\bibinfo {year} {2010})}\BibitemShut {NoStop}%
\bibitem [{\citenamefont {Sorre}\ \emph {et~al.}(2012)\citenamefont {Sorre}, \citenamefont {Callan-Jones}, \citenamefont {Manzi}, \citenamefont {Goud}, \citenamefont {Prost}, \citenamefont {Bassereau},\ and\ \citenamefont {Roux}}]{sorre2012nature}%
  \BibitemOpen
  \bibfield  {author} {\bibinfo {author} {\bibfnamefont {B.}~\bibnamefont {Sorre}}, \bibinfo {author} {\bibfnamefont {A.}~\bibnamefont {Callan-Jones}}, \bibinfo {author} {\bibfnamefont {J.}~\bibnamefont {Manzi}}, \bibinfo {author} {\bibfnamefont {B.}~\bibnamefont {Goud}}, \bibinfo {author} {\bibfnamefont {J.}~\bibnamefont {Prost}}, \bibinfo {author} {\bibfnamefont {P.}~\bibnamefont {Bassereau}},\ and\ \bibinfo {author} {\bibfnamefont {A.}~\bibnamefont {Roux}},\ }\href {https://doi.org/10.1073/pnas.1103594108} {\bibfield  {journal} {\bibinfo  {journal} {Proc. Natl. Acad. Sci. U.S.A.}\ }\textbf {\bibinfo {volume} {109}},\ \bibinfo {pages} {173} (\bibinfo {year} {2012})}\BibitemShut {NoStop}%
\bibitem [{\citenamefont {Beltr{\'a}n-Heredia}\ \emph {et~al.}(2019)\citenamefont {Beltr{\'a}n-Heredia}, \citenamefont {Tsai}, \citenamefont {Salinas-Almaguer}, \citenamefont {Cao}, \citenamefont {Bassereau},\ and\ \citenamefont {Monroy}}]{beltran2019membrane}%
  \BibitemOpen
  \bibfield  {author} {\bibinfo {author} {\bibfnamefont {E.}~\bibnamefont {Beltr{\'a}n-Heredia}}, \bibinfo {author} {\bibfnamefont {F.~C.}\ \bibnamefont {Tsai}}, \bibinfo {author} {\bibfnamefont {S.}~\bibnamefont {Salinas-Almaguer}}, \bibinfo {author} {\bibfnamefont {F.~J.}\ \bibnamefont {Cao}}, \bibinfo {author} {\bibfnamefont {P.}~\bibnamefont {Bassereau}},\ and\ \bibinfo {author} {\bibfnamefont {F.}~\bibnamefont {Monroy}},\ }\href {https://doi.org/10.1038/s42003-019-0471-x} {\bibfield  {journal} {\bibinfo  {journal} {Commun. Biol.}\ }\textbf {\bibinfo {volume} {2}},\ \bibinfo {pages} {225} (\bibinfo {year} {2019})}\BibitemShut {NoStop}%
\bibitem [{\citenamefont {Tian}\ and\ \citenamefont {Baumgart}(2009)}]{tian2009sorting}%
  \BibitemOpen
  \bibfield  {author} {\bibinfo {author} {\bibfnamefont {A.}~\bibnamefont {Tian}}\ and\ \bibinfo {author} {\bibfnamefont {T.}~\bibnamefont {Baumgart}},\ }\href {https://doi.org/10.1016/j.bpj.2008.11.067} {\bibfield  {journal} {\bibinfo  {journal} {Biophys. J.}\ }\textbf {\bibinfo {volume} {96}},\ \bibinfo {pages} {2676} (\bibinfo {year} {2009})}\BibitemShut {NoStop}%
\bibitem [{\citenamefont {Kamal}\ \emph {et~al.}(2009)\citenamefont {Kamal}, \citenamefont {Mills}, \citenamefont {Grzybek},\ and\ \citenamefont {Howard}}]{kamal2009measurement}%
  \BibitemOpen
  \bibfield  {author} {\bibinfo {author} {\bibfnamefont {M.~M.}\ \bibnamefont {Kamal}}, \bibinfo {author} {\bibfnamefont {D.}~\bibnamefont {Mills}}, \bibinfo {author} {\bibfnamefont {M.}~\bibnamefont {Grzybek}},\ and\ \bibinfo {author} {\bibfnamefont {J.}~\bibnamefont {Howard}},\ }\href {https://doi.org/10.1073/pnas.0907354106} {\bibfield  {journal} {\bibinfo  {journal} {Proc. Natl. Acad. Sci. U.S.A.}\ }\textbf {\bibinfo {volume} {106}},\ \bibinfo {pages} {22245} (\bibinfo {year} {2009})}\BibitemShut {NoStop}%
\bibitem [{\citenamefont {Sorre}\ \emph {et~al.}(2009)\citenamefont {Sorre}, \citenamefont {Callan-Jones}, \citenamefont {Manneville}, \citenamefont {Nassoy}, \citenamefont {Joanny}, \citenamefont {Prost}, \citenamefont {Goud},\ and\ \citenamefont {Bassereau}}]{sorre2009curvature}%
  \BibitemOpen
  \bibfield  {author} {\bibinfo {author} {\bibfnamefont {B.}~\bibnamefont {Sorre}}, \bibinfo {author} {\bibfnamefont {A.}~\bibnamefont {Callan-Jones}}, \bibinfo {author} {\bibfnamefont {J.~B.}\ \bibnamefont {Manneville}}, \bibinfo {author} {\bibfnamefont {P.}~\bibnamefont {Nassoy}}, \bibinfo {author} {\bibfnamefont {J.~F.}\ \bibnamefont {Joanny}}, \bibinfo {author} {\bibfnamefont {J.}~\bibnamefont {Prost}}, \bibinfo {author} {\bibfnamefont {B.}~\bibnamefont {Goud}},\ and\ \bibinfo {author} {\bibfnamefont {P.}~\bibnamefont {Bassereau}},\ }\href {https://doi.org/10.1073/pnas.0811243106} {\bibfield  {journal} {\bibinfo  {journal} {Proc. Natl. Acad. Sci. U.S.A.}\ }\textbf {\bibinfo {volume} {106}},\ \bibinfo {pages} {5622} (\bibinfo {year} {2009})}\BibitemShut {NoStop}%
\bibitem [{\citenamefont {Parthasarathy}\ \emph {et~al.}(2006)\citenamefont {Parthasarathy}, \citenamefont {Yu},\ and\ \citenamefont {Groves}}]{parthasarathy2006curvature}%
  \BibitemOpen
  \bibfield  {author} {\bibinfo {author} {\bibfnamefont {R.}~\bibnamefont {Parthasarathy}}, \bibinfo {author} {\bibfnamefont {C.~H.}\ \bibnamefont {Yu}},\ and\ \bibinfo {author} {\bibfnamefont {J.~T.}\ \bibnamefont {Groves}},\ }\href {https://doi.org/10.1021/la060390o} {\bibfield  {journal} {\bibinfo  {journal} {Langmuir}\ }\textbf {\bibinfo {volume} {22}},\ \bibinfo {pages} {5095} (\bibinfo {year} {2006})}\BibitemShut {NoStop}%
\bibitem [{\citenamefont {Yoon}\ \emph {et~al.}(2006)\citenamefont {Yoon}, \citenamefont {Jeong}, \citenamefont {Lee}, \citenamefont {Kim}, \citenamefont {Choi}, \citenamefont {Kim}, \citenamefont {Kim},\ and\ \citenamefont {Lee}}]{yoon2006topographic}%
  \BibitemOpen
  \bibfield  {author} {\bibinfo {author} {\bibfnamefont {T.~Y.}\ \bibnamefont {Yoon}}, \bibinfo {author} {\bibfnamefont {C.}~\bibnamefont {Jeong}}, \bibinfo {author} {\bibfnamefont {S.~W.}\ \bibnamefont {Lee}}, \bibinfo {author} {\bibfnamefont {J.~H.}\ \bibnamefont {Kim}}, \bibinfo {author} {\bibfnamefont {M.~C.}\ \bibnamefont {Choi}}, \bibinfo {author} {\bibfnamefont {S.~J.}\ \bibnamefont {Kim}}, \bibinfo {author} {\bibfnamefont {M.~W.}\ \bibnamefont {Kim}},\ and\ \bibinfo {author} {\bibfnamefont {S.~D.}\ \bibnamefont {Lee}},\ }\href {https://doi.org/10.1038/nmat1618} {\bibfield  {journal} {\bibinfo  {journal} {Nat. Mater.}\ }\textbf {\bibinfo {volume} {5}},\ \bibinfo {pages} {281} (\bibinfo {year} {2006})}\BibitemShut {NoStop}%
\bibitem [{\citenamefont {Hsieh}\ \emph {et~al.}(2012)\citenamefont {Hsieh}, \citenamefont {Hsu}, \citenamefont {Capraro}, \citenamefont {Wu}, \citenamefont {Chen}, \citenamefont {Yang},\ and\ \citenamefont {Baumgart}}]{hsieh2012curvature}%
  \BibitemOpen
  \bibfield  {author} {\bibinfo {author} {\bibfnamefont {W.~T.}\ \bibnamefont {Hsieh}}, \bibinfo {author} {\bibfnamefont {C.~J.}\ \bibnamefont {Hsu}}, \bibinfo {author} {\bibfnamefont {B.~R.}\ \bibnamefont {Capraro}}, \bibinfo {author} {\bibfnamefont {T.}~\bibnamefont {Wu}}, \bibinfo {author} {\bibfnamefont {C.~M.}\ \bibnamefont {Chen}}, \bibinfo {author} {\bibfnamefont {S.}~\bibnamefont {Yang}},\ and\ \bibinfo {author} {\bibfnamefont {T.}~\bibnamefont {Baumgart}},\ }\href {https://doi.org/10.1021/la302205b} {\bibfield  {journal} {\bibinfo  {journal} {Langmuir}\ }\textbf {\bibinfo {volume} {28}},\ \bibinfo {pages} {12838} (\bibinfo {year} {2012})}\BibitemShut {NoStop}%
\bibitem [{\citenamefont {Black}\ \emph {et~al.}(2014)\citenamefont {Black}, \citenamefont {Cheney}, \citenamefont {Campbell},\ and\ \citenamefont {Knowles}}]{black2014membrane}%
  \BibitemOpen
  \bibfield  {author} {\bibinfo {author} {\bibfnamefont {J.~C.}\ \bibnamefont {Black}}, \bibinfo {author} {\bibfnamefont {P.~P.}\ \bibnamefont {Cheney}}, \bibinfo {author} {\bibfnamefont {T.}~\bibnamefont {Campbell}},\ and\ \bibinfo {author} {\bibfnamefont {M.~K.}\ \bibnamefont {Knowles}},\ }\href {https://doi.org/10.1039/c3sm52522h} {\bibfield  {journal} {\bibinfo  {journal} {Soft Matter}\ }\textbf {\bibinfo {volume} {10}},\ \bibinfo {pages} {2016} (\bibinfo {year} {2014})}\BibitemShut {NoStop}%
\bibitem [{\citenamefont {Renner}\ and\ \citenamefont {Weibel}(2011)}]{renner2011cardiolipin}%
  \BibitemOpen
  \bibfield  {author} {\bibinfo {author} {\bibfnamefont {L.~D.}\ \bibnamefont {Renner}}\ and\ \bibinfo {author} {\bibfnamefont {D.~B.}\ \bibnamefont {Weibel}},\ }\href {https://doi.org/10.1073/pnas.1015757108} {\bibfield  {journal} {\bibinfo  {journal} {Proc. Natl. Acad. Sci. U.S.A.}\ }\textbf {\bibinfo {volume} {108}},\ \bibinfo {pages} {6264} (\bibinfo {year} {2011})}\BibitemShut {NoStop}%
\bibitem [{\citenamefont {Salbreux}\ \emph {et~al.}(2012)\citenamefont {Salbreux}, \citenamefont {Charras},\ and\ \citenamefont {Paluch}}]{salbreux2012actin}%
  \BibitemOpen
  \bibfield  {author} {\bibinfo {author} {\bibfnamefont {G.}~\bibnamefont {Salbreux}}, \bibinfo {author} {\bibfnamefont {G.}~\bibnamefont {Charras}},\ and\ \bibinfo {author} {\bibfnamefont {E.}~\bibnamefont {Paluch}},\ }\href {https://doi.org/10.1016/j.tcb.2012.07.001 External Link} {\bibfield  {journal} {\bibinfo  {journal} {Trends Cell Biol.}\ }\textbf {\bibinfo {volume} {22}},\ \bibinfo {pages} {536} (\bibinfo {year} {2012})}\BibitemShut {NoStop}%
\bibitem [{\citenamefont {Frank}\ \emph {et~al.}(2019)\citenamefont {Frank}, \citenamefont {Guven}, \citenamefont {Kardar},\ and\ \citenamefont {Shackleton}}]{frank2019pinning}%
  \BibitemOpen
  \bibfield  {author} {\bibinfo {author} {\bibfnamefont {J.~R.}\ \bibnamefont {Frank}}, \bibinfo {author} {\bibfnamefont {J.}~\bibnamefont {Guven}}, \bibinfo {author} {\bibfnamefont {M.}~\bibnamefont {Kardar}},\ and\ \bibinfo {author} {\bibfnamefont {H.}~\bibnamefont {Shackleton}},\ }\href {https://doi.org/10.1209/0295-5075/127/48001} {\bibfield  {journal} {\bibinfo  {journal} {Europhys. Lett.}\ }\textbf {\bibinfo {volume} {127}},\ \bibinfo {pages} {48001} (\bibinfo {year} {2019})}\BibitemShut {NoStop}%
\bibitem [{\citenamefont {Bowick}\ and\ \citenamefont {Giomi}(2009)}]{bowick2009two}%
  \BibitemOpen
  \bibfield  {author} {\bibinfo {author} {\bibfnamefont {M.~J.}\ \bibnamefont {Bowick}}\ and\ \bibinfo {author} {\bibfnamefont {L.}~\bibnamefont {Giomi}},\ }\href {https://doi.org/10.1080/00018730903043166} {\bibfield  {journal} {\bibinfo  {journal} {Adv. Phys.}\ }\textbf {\bibinfo {volume} {58}},\ \bibinfo {pages} {449} (\bibinfo {year} {2009})}\BibitemShut {NoStop}%
\bibitem [{\citenamefont {Deserno}(2015)}]{deserno2015fluid}%
  \BibitemOpen
  \bibfield  {author} {\bibinfo {author} {\bibfnamefont {M.}~\bibnamefont {Deserno}},\ }\href {https://doi.org/10.1016/j.chemphyslip.2014.05.001} {\bibfield  {journal} {\bibinfo  {journal} {Chem. Phys. Lipids}\ }\textbf {\bibinfo {volume} {185}},\ \bibinfo {pages} {11} (\bibinfo {year} {2015})}\BibitemShut {NoStop}%
\bibitem [{\citenamefont {Moriel}\ and\ \citenamefont {Stone}()}]{SM}%
  \BibitemOpen
  \bibfield  {author} {\bibinfo {author} {\bibfnamefont {A.}~\bibnamefont {Moriel}}\ and\ \bibinfo {author} {\bibfnamefont {H.~A.}\ \bibnamefont {Stone}},\ }\href@noop {} {\bibinfo {title} {Supplemental material}}\BibitemShut {NoStop}%
\bibitem [{\citenamefont {Jenkins}(1977{\natexlab{a}})}]{jenkins1977equations}%
  \BibitemOpen
  \bibfield  {author} {\bibinfo {author} {\bibfnamefont {J.~T.}\ \bibnamefont {Jenkins}},\ }\href@noop {} {\bibfield  {journal} {\bibinfo  {journal} {SIAM J. Appl. Math.}\ }\textbf {\bibinfo {volume} {32}},\ \bibinfo {pages} {755} (\bibinfo {year} {1977}{\natexlab{a}})}\BibitemShut {NoStop}%
\bibitem [{\citenamefont {Jenkins}(1977{\natexlab{b}})}]{jenkins1977static}%
  \BibitemOpen
  \bibfield  {author} {\bibinfo {author} {\bibfnamefont {J.~T.}\ \bibnamefont {Jenkins}},\ }\href {https://doi.org/10.1007/BF00275981} {\bibfield  {journal} {\bibinfo  {journal} {J. Math. Biol.}\ }\textbf {\bibinfo {volume} {4}},\ \bibinfo {pages} {149} (\bibinfo {year} {1977}{\natexlab{b}})}\BibitemShut {NoStop}%
\bibitem [{\citenamefont {Zhong-Can}\ and\ \citenamefont {Helfrich}(1987)}]{zhong1987instability}%
  \BibitemOpen
  \bibfield  {author} {\bibinfo {author} {\bibfnamefont {O.~Y.}\ \bibnamefont {Zhong-Can}}\ and\ \bibinfo {author} {\bibfnamefont {W.}~\bibnamefont {Helfrich}},\ }\href {https://doi.org/10.1103/PhysRevLett.59.2486} {\bibfield  {journal} {\bibinfo  {journal} {Phys. Rev. Lett.}\ }\textbf {\bibinfo {volume} {59}},\ \bibinfo {pages} {2486} (\bibinfo {year} {1987})}\BibitemShut {NoStop}%
\bibitem [{\citenamefont {Zhong-Can}\ and\ \citenamefont {Helfrich}(1989)}]{zhong1989bending}%
  \BibitemOpen
  \bibfield  {author} {\bibinfo {author} {\bibfnamefont {O.~Y.}\ \bibnamefont {Zhong-Can}}\ and\ \bibinfo {author} {\bibfnamefont {W.}~\bibnamefont {Helfrich}},\ }\href {https://doi.org/10.1103/PhysRevA.39.5280} {\bibfield  {journal} {\bibinfo  {journal} {Phys. Rev. A}\ }\textbf {\bibinfo {volume} {39}},\ \bibinfo {pages} {5280} (\bibinfo {year} {1989})}\BibitemShut {NoStop}%
\bibitem [{\citenamefont {Capovilla}\ \emph {et~al.}(2003)\citenamefont {Capovilla}, \citenamefont {Guven},\ and\ \citenamefont {Santiago}}]{capovilla2003deformations}%
  \BibitemOpen
  \bibfield  {author} {\bibinfo {author} {\bibfnamefont {R.}~\bibnamefont {Capovilla}}, \bibinfo {author} {\bibfnamefont {J.}~\bibnamefont {Guven}},\ and\ \bibinfo {author} {\bibfnamefont {J.~A.}\ \bibnamefont {Santiago}},\ }\href {https://doi.org/10.1088/0305-4470/36/23/301} {\bibfield  {journal} {\bibinfo  {journal} {J. Phys. A: Math. Gen.}\ }\textbf {\bibinfo {volume} {36}},\ \bibinfo {pages} {6281} (\bibinfo {year} {2003})}\BibitemShut {NoStop}%
\bibitem [{\citenamefont {Spivak}(1970)}]{spivak1970comprehensive}%
  \BibitemOpen
  \bibfield  {author} {\bibinfo {author} {\bibfnamefont {M.~D.}\ \bibnamefont {Spivak}},\ }\href@noop {} {\bibfield  {journal} {\bibinfo  {journal} {(No Title)}\ } (\bibinfo {year} {1970})}\BibitemShut {NoStop}%
\bibitem [{\citenamefont {Hersch}(1970)}]{hersch1970quatre}%
  \BibitemOpen
  \bibfield  {author} {\bibinfo {author} {\bibfnamefont {J.}~\bibnamefont {Hersch}},\ }\href@noop {} {\bibfield  {journal} {\bibinfo  {journal} {C. R. Acad. Sci. Paris Sér. A-B}\ }\textbf {\bibinfo {volume} {270}},\ \bibinfo {pages} {A1645} (\bibinfo {year} {1970})}\BibitemShut {NoStop}%
\bibitem [{\citenamefont {Nishide}\ and\ \citenamefont {Ishihara}(2025)}]{nishide2025oscillatory}%
  \BibitemOpen
  \bibfield  {author} {\bibinfo {author} {\bibfnamefont {R.}~\bibnamefont {Nishide}}\ and\ \bibinfo {author} {\bibfnamefont {S.}~\bibnamefont {Ishihara}},\ }\href {https://doi.org/10.1103/PhysRevE.111.L022202} {\bibfield  {journal} {\bibinfo  {journal} {Phys. Rev. E}\ }\textbf {\bibinfo {volume} {111}},\ \bibinfo {pages} {L022202} (\bibinfo {year} {2025})}\BibitemShut {NoStop}%
\bibitem [{\citenamefont {Nishide}\ and\ \citenamefont {Ishihara}(2022)}]{nishide2022pattern}%
  \BibitemOpen
  \bibfield  {author} {\bibinfo {author} {\bibfnamefont {R.}~\bibnamefont {Nishide}}\ and\ \bibinfo {author} {\bibfnamefont {S.}~\bibnamefont {Ishihara}},\ }\href {https://doi.org/10.1103/PhysRevLett.128.224101} {\bibfield  {journal} {\bibinfo  {journal} {Phys. Rev. Lett.}\ }\textbf {\bibinfo {volume} {128}},\ \bibinfo {pages} {224101} (\bibinfo {year} {2022})}\BibitemShut {NoStop}%
\bibitem [{\citenamefont {Tamemoto}\ and\ \citenamefont {Noguchi}(2021)}]{tamemoto2021reaction}%
  \BibitemOpen
  \bibfield  {author} {\bibinfo {author} {\bibfnamefont {N.}~\bibnamefont {Tamemoto}}\ and\ \bibinfo {author} {\bibfnamefont {H.}~\bibnamefont {Noguchi}},\ }\href {https://doi.org/10.1039/d1sm00540e} {\bibfield  {journal} {\bibinfo  {journal} {Soft Matter}\ }\textbf {\bibinfo {volume} {17}},\ \bibinfo {pages} {6589} (\bibinfo {year} {2021})}\BibitemShut {NoStop}%
\bibitem [{\citenamefont {Noguchi}(2023)}]{noguchi2023disappearance}%
  \BibitemOpen
  \bibfield  {author} {\bibinfo {author} {\bibfnamefont {H.}~\bibnamefont {Noguchi}},\ }\href {https://doi.org/10.1038/s41598-023-33376-9} {\bibfield  {journal} {\bibinfo  {journal} {Sci. Rep.}\ }\textbf {\bibinfo {volume} {13}},\ \bibinfo {pages} {6207} (\bibinfo {year} {2023})}\BibitemShut {NoStop}%
\bibitem [{\citenamefont {Singh}\ \emph {et~al.}(2022)\citenamefont {Singh}, \citenamefont {Leadbetter},\ and\ \citenamefont {Camley}}]{singh2022sensing}%
  \BibitemOpen
  \bibfield  {author} {\bibinfo {author} {\bibfnamefont {A.~R.}\ \bibnamefont {Singh}}, \bibinfo {author} {\bibfnamefont {T.}~\bibnamefont {Leadbetter}},\ and\ \bibinfo {author} {\bibfnamefont {B.~A.}\ \bibnamefont {Camley}},\ }\href {https://doi.org/10.1073/pnas.2121302119} {\bibfield  {journal} {\bibinfo  {journal} {Proc. Natl. Acad. Sci. U.S.A.}\ }\textbf {\bibinfo {volume} {119}},\ \bibinfo {pages} {e2121302119} (\bibinfo {year} {2022})}\BibitemShut {NoStop}%
\bibitem [{\citenamefont {Orlandini}\ \emph {et~al.}(2013)\citenamefont {Orlandini}, \citenamefont {Marenduzzo},\ and\ \citenamefont {Goryachev}}]{orlandini2013domain}%
  \BibitemOpen
  \bibfield  {author} {\bibinfo {author} {\bibfnamefont {E.}~\bibnamefont {Orlandini}}, \bibinfo {author} {\bibfnamefont {D.}~\bibnamefont {Marenduzzo}},\ and\ \bibinfo {author} {\bibfnamefont {A.~B.}\ \bibnamefont {Goryachev}},\ }\href {https://doi.org/10.1039/c3sm50650a} {\bibfield  {journal} {\bibinfo  {journal} {Soft Matter}\ }\textbf {\bibinfo {volume} {9}},\ \bibinfo {pages} {9311} (\bibinfo {year} {2013})}\BibitemShut {NoStop}%
\bibitem [{\citenamefont {Ginzburg}\ and\ \citenamefont {Landau}(2009)}]{ginzburg2009theory}%
  \BibitemOpen
  \bibfield  {author} {\bibinfo {author} {\bibfnamefont {V.~L.}\ \bibnamefont {Ginzburg}}\ and\ \bibinfo {author} {\bibfnamefont {L.~D.}\ \bibnamefont {Landau}},\ }in\ \href {https://doi.org/10.1007/978-3-540-68008-6_4} {\emph {\bibinfo {booktitle} {On superconductivity and superfluidity: a scientific autobiography}}}\ (\bibinfo  {publisher} {Springer},\ \bibinfo {year} {2009})\ pp.\ \bibinfo {pages} {113--137}\BibitemShut {NoStop}%
\bibitem [{\citenamefont {Kardar}(2007)}]{kardar2007statistical}%
  \BibitemOpen
  \bibfield  {author} {\bibinfo {author} {\bibfnamefont {M.}~\bibnamefont {Kardar}},\ }\href@noop {} {\emph {\bibinfo {title} {Statistical physics of fields}}}\ (\bibinfo  {publisher} {Cambridge University Press},\ \bibinfo {year} {2007})\BibitemShut {NoStop}%
\bibitem [{\citenamefont {Safran}(2018)}]{safran2018statistical}%
  \BibitemOpen
  \bibfield  {author} {\bibinfo {author} {\bibfnamefont {S.}~\bibnamefont {Safran}},\ }\href@noop {} {\emph {\bibinfo {title} {Statistical thermodynamics of surfaces, interfaces, and membranes}}}\ (\bibinfo  {publisher} {CRC Press},\ \bibinfo {year} {2018})\BibitemShut {NoStop}%
\bibitem [{\citenamefont {Cahn}(1961)}]{cahn1961spinodal}%
  \BibitemOpen
  \bibfield  {author} {\bibinfo {author} {\bibfnamefont {J.~W.}\ \bibnamefont {Cahn}},\ }\href {https://doi.org/10.1016/0001-6160(61)90182-1} {\bibfield  {journal} {\bibinfo  {journal} {Acta Metall.}\ }\textbf {\bibinfo {volume} {9}},\ \bibinfo {pages} {795} (\bibinfo {year} {1961})}\BibitemShut {NoStop}%
\bibitem [{\citenamefont {Cahn}\ and\ \citenamefont {Hilliard}(1958)}]{cahn1958interfacial}%
  \BibitemOpen
  \bibfield  {author} {\bibinfo {author} {\bibfnamefont {J.~W.}\ \bibnamefont {Cahn}}\ and\ \bibinfo {author} {\bibfnamefont {J.~E.}\ \bibnamefont {Hilliard}},\ }\href {https://doi.org/10.1063/1.1744102} {\bibfield  {journal} {\bibinfo  {journal} {J. Chem. Phys.}\ }\textbf {\bibinfo {volume} {28}},\ \bibinfo {pages} {258} (\bibinfo {year} {1958})}\BibitemShut {NoStop}%
\bibitem [{\citenamefont {G{\'o}mez}\ \emph {et~al.}(2015)\citenamefont {G{\'o}mez}, \citenamefont {Garc{\'\i}a}, \citenamefont {Vitelli}, \citenamefont {Lorenzana},\ and\ \citenamefont {Vega}}]{gomez2015phase}%
  \BibitemOpen
  \bibfield  {author} {\bibinfo {author} {\bibfnamefont {L.~R.}\ \bibnamefont {G{\'o}mez}}, \bibinfo {author} {\bibfnamefont {N.~A.}\ \bibnamefont {Garc{\'\i}a}}, \bibinfo {author} {\bibfnamefont {V.}~\bibnamefont {Vitelli}}, \bibinfo {author} {\bibfnamefont {J.}~\bibnamefont {Lorenzana}},\ and\ \bibinfo {author} {\bibfnamefont {D.~A.}\ \bibnamefont {Vega}},\ }\href {https://doi.org/10.1038/ncomms7856} {\bibfield  {journal} {\bibinfo  {journal} {Nat. Commun.}\ }\textbf {\bibinfo {volume} {6}},\ \bibinfo {pages} {6856} (\bibinfo {year} {2015})}\BibitemShut {NoStop}%
\end{thebibliography}

\begin{thebibliography}{14}%
\makeatletter
\providecommand \@ifxundefined [1]{%
 \@ifx{#1\undefined}
}%
\providecommand \@ifnum [1]{%
 \ifnum #1\expandafter \@firstoftwo
 \else \expandafter \@secondoftwo
 \fi
}%
\providecommand \@ifx [1]{%
 \ifx #1\expandafter \@firstoftwo
 \else \expandafter \@secondoftwo
 \fi
}%
\providecommand \natexlab [1]{#1}%
\providecommand \enquote  [1]{``#1''}%
\providecommand \bibnamefont  [1]{#1}%
\providecommand \bibfnamefont [1]{#1}%
\providecommand \citenamefont [1]{#1}%
\providecommand \href@noop [0]{\@secondoftwo}%
\providecommand \href [0]{\begingroup \@sanitize@url \@href}%
\providecommand \@href[1]{\@@startlink{#1}\@@href}%
\providecommand \@@href[1]{\endgroup#1\@@endlink}%
\providecommand \@sanitize@url [0]{\catcode `\\12\catcode `\$12\catcode `\&12\catcode `\#12\catcode `\^12\catcode `\_12\catcode `\%12\relax}%
\providecommand \@@startlink[1]{}%
\providecommand \@@endlink[0]{}%
\providecommand \url  [0]{\begingroup\@sanitize@url \@url }%
\providecommand \@url [1]{\endgroup\@href {#1}{\urlprefix }}%
\providecommand \urlprefix  [0]{URL }%
\providecommand \Eprint [0]{\href }%
\providecommand \doibase [0]{https://doi.org/}%
\providecommand \selectlanguage [0]{\@gobble}%
\providecommand \bibinfo  [0]{\@secondoftwo}%
\providecommand \bibfield  [0]{\@secondoftwo}%
\providecommand \translation [1]{[#1]}%
\providecommand \BibitemOpen [0]{}%
\providecommand \bibitemStop [0]{}%
\providecommand \bibitemNoStop [0]{.\EOS\space}%
\providecommand \EOS [0]{\spacefactor3000\relax}%
\providecommand \BibitemShut  [1]{\csname bibitem#1\endcsname}%
\let\auto@bib@innerbib\@empty
\bibitem [{\citenamefont {Deserno}(2015)}]{Sdeserno2015fluid}%
  \BibitemOpen
  \bibfield  {author} {\bibinfo {author} {\bibfnamefont {M.}~\bibnamefont {Deserno}},\ }\href {https://doi.org/10.1016/j.chemphyslip.2014.05.001} {\bibfield  {journal} {\bibinfo  {journal} {Chem. Phys. Lipids}\ }\textbf {\bibinfo {volume} {185}},\ \bibinfo {pages} {11} (\bibinfo {year} {2015})}\BibitemShut {NoStop}%
\bibitem [{\citenamefont {Kardar}(2007)}]{Skardar2007statistical}%
  \BibitemOpen
  \bibfield  {author} {\bibinfo {author} {\bibfnamefont {M.}~\bibnamefont {Kardar}},\ }\href@noop {} {\emph {\bibinfo {title} {Statistical physics of fields}}}\ (\bibinfo  {publisher} {Cambridge University Press},\ \bibinfo {year} {2007})\BibitemShut {NoStop}%
\bibitem [{\citenamefont {Safran}(2018)}]{Ssafran2018statistical}%
  \BibitemOpen
  \bibfield  {author} {\bibinfo {author} {\bibfnamefont {S.}~\bibnamefont {Safran}},\ }\href@noop {} {\emph {\bibinfo {title} {Statistical thermodynamics of surfaces, interfaces, and membranes}}}\ (\bibinfo  {publisher} {CRC Press},\ \bibinfo {year} {2018})\BibitemShut {NoStop}%
\bibitem [{\citenamefont {Cheeger}(2015)}]{Scheeger2015lower}%
  \BibitemOpen
  \bibfield  {author} {\bibinfo {author} {\bibfnamefont {J.}~\bibnamefont {Cheeger}},\ }in\ \href@noop {} {\emph {\bibinfo {booktitle} {Problems in analysis}}}\ (\bibinfo  {publisher} {Princeton University Press},\ \bibinfo {year} {2015})\ pp.\ \bibinfo {pages} {195--200}\BibitemShut {NoStop}%
\bibitem [{\citenamefont {Sullivan}\ and\ \citenamefont {Kaszynski}(2019)}]{Ssullivan2019pyvista}%
  \BibitemOpen
  \bibfield  {author} {\bibinfo {author} {\bibfnamefont {B.}~\bibnamefont {Sullivan}}\ and\ \bibinfo {author} {\bibfnamefont {A.}~\bibnamefont {Kaszynski}},\ }\href {https://doi.org/10.21105/joss.01450} {\bibfield  {journal} {\bibinfo  {journal} {Journal of Open Source Software}\ }\textbf {\bibinfo {volume} {4}},\ \bibinfo {pages} {1450} (\bibinfo {year} {2019})}\BibitemShut {NoStop}%
\bibitem [{\citenamefont {Virtanen}\ \emph {et~al.}(2020)\citenamefont {Virtanen}, \citenamefont {Gommers}, \citenamefont {Oliphant}, \citenamefont {Haberland}, \citenamefont {Reddy}, \citenamefont {Cournapeau}, \citenamefont {Burovski}, \citenamefont {Peterson}, \citenamefont {Weckesser}, \citenamefont {Bright}, \citenamefont {{van der Walt}}, \citenamefont {Brett}, \citenamefont {Wilson}, \citenamefont {Millman}, \citenamefont {Mayorov}, \citenamefont {Nelson}, \citenamefont {Jones}, \citenamefont {Kern}, \citenamefont {Larson}, \citenamefont {Carey}, \citenamefont {Polat}, \citenamefont {Feng}, \citenamefont {Moore}, \citenamefont {{VanderPlas}}, \citenamefont {Laxalde}, \citenamefont {Perktold}, \citenamefont {Cimrman}, \citenamefont {Henriksen}, \citenamefont {Quintero}, \citenamefont {Harris}, \citenamefont {Archibald}, \citenamefont {Ribeiro}, \citenamefont {Pedregosa}, \citenamefont {{van Mulbregt}},\ and\ \citenamefont {{SciPy 1.0 Contributors}}}]{S2020SciPy-NMeth}%
  \BibitemOpen
  \bibfield  {author} {\bibinfo {author} {\bibfnamefont {P.}~\bibnamefont {Virtanen}}, \bibinfo {author} {\bibfnamefont {R.}~\bibnamefont {Gommers}}, \bibinfo {author} {\bibfnamefont {T.~E.}\ \bibnamefont {Oliphant}}, \bibinfo {author} {\bibfnamefont {M.}~\bibnamefont {Haberland}}, \bibinfo {author} {\bibfnamefont {T.}~\bibnamefont {Reddy}}, \bibinfo {author} {\bibfnamefont {D.}~\bibnamefont {Cournapeau}}, \bibinfo {author} {\bibfnamefont {E.}~\bibnamefont {Burovski}}, \bibinfo {author} {\bibfnamefont {P.}~\bibnamefont {Peterson}}, \bibinfo {author} {\bibfnamefont {W.}~\bibnamefont {Weckesser}}, \bibinfo {author} {\bibfnamefont {J.}~\bibnamefont {Bright}}, \bibinfo {author} {\bibfnamefont {S.~J.}\ \bibnamefont {{van der Walt}}}, \bibinfo {author} {\bibfnamefont {M.}~\bibnamefont {Brett}}, \bibinfo {author} {\bibfnamefont {J.}~\bibnamefont {Wilson}}, \bibinfo {author} {\bibfnamefont {K.~J.}\ \bibnamefont {Millman}}, \bibinfo {author} {\bibfnamefont {N.}~\bibnamefont {Mayorov}}, \bibinfo {author} {\bibfnamefont
  {A.~R.~J.}\ \bibnamefont {Nelson}}, \bibinfo {author} {\bibfnamefont {E.}~\bibnamefont {Jones}}, \bibinfo {author} {\bibfnamefont {R.}~\bibnamefont {Kern}}, \bibinfo {author} {\bibfnamefont {E.}~\bibnamefont {Larson}}, \bibinfo {author} {\bibfnamefont {C.~J.}\ \bibnamefont {Carey}}, \bibinfo {author} {\bibfnamefont {{\.I}.}~\bibnamefont {Polat}}, \bibinfo {author} {\bibfnamefont {Y.}~\bibnamefont {Feng}}, \bibinfo {author} {\bibfnamefont {E.~W.}\ \bibnamefont {Moore}}, \bibinfo {author} {\bibfnamefont {J.}~\bibnamefont {{VanderPlas}}}, \bibinfo {author} {\bibfnamefont {D.}~\bibnamefont {Laxalde}}, \bibinfo {author} {\bibfnamefont {J.}~\bibnamefont {Perktold}}, \bibinfo {author} {\bibfnamefont {R.}~\bibnamefont {Cimrman}}, \bibinfo {author} {\bibfnamefont {I.}~\bibnamefont {Henriksen}}, \bibinfo {author} {\bibfnamefont {E.~A.}\ \bibnamefont {Quintero}}, \bibinfo {author} {\bibfnamefont {C.~R.}\ \bibnamefont {Harris}}, \bibinfo {author} {\bibfnamefont {A.~M.}\ \bibnamefont {Archibald}}, \bibinfo {author}
  {\bibfnamefont {A.~H.}\ \bibnamefont {Ribeiro}}, \bibinfo {author} {\bibfnamefont {F.}~\bibnamefont {Pedregosa}}, \bibinfo {author} {\bibfnamefont {P.}~\bibnamefont {{van Mulbregt}}},\ and\ \bibinfo {author} {\bibnamefont {{SciPy 1.0 Contributors}}},\ }\href {https://doi.org/10.1038/s41592-019-0686-2} {\bibfield  {journal} {\bibinfo  {journal} {Nat. Methods}\ }\textbf {\bibinfo {volume} {17}},\ \bibinfo {pages} {261} (\bibinfo {year} {2020})}\BibitemShut {NoStop}%
\bibitem [{\citenamefont {Reuter}\ \emph {et~al.}(2006)\citenamefont {Reuter}, \citenamefont {Wolter},\ and\ \citenamefont {Peinecke}}]{Sreuter2006laplace}%
  \BibitemOpen
  \bibfield  {author} {\bibinfo {author} {\bibfnamefont {M.}~\bibnamefont {Reuter}}, \bibinfo {author} {\bibfnamefont {F.~E.}\ \bibnamefont {Wolter}},\ and\ \bibinfo {author} {\bibfnamefont {N.}~\bibnamefont {Peinecke}},\ }\href {https://doi.org/10.1016/j.cad.2005.10.011} {\bibfield  {journal} {\bibinfo  {journal} {Computer-Aided Design}\ }\textbf {\bibinfo {volume} {38}},\ \bibinfo {pages} {342} (\bibinfo {year} {2006})}\BibitemShut {NoStop}%
\bibitem [{\citenamefont {Baratta}\ \emph {et~al.}(2023)\citenamefont {Baratta}, \citenamefont {Dean}, \citenamefont {Dokken}, \citenamefont {Habera}, \citenamefont {Hale}, \citenamefont {Richardson}, \citenamefont {Rognes}, \citenamefont {Scroggs}, \citenamefont {Sime},\ and\ \citenamefont {Wells}}]{Sbaratta2023dolfinx}%
  \BibitemOpen
  \bibfield  {author} {\bibinfo {author} {\bibfnamefont {I.~A.}\ \bibnamefont {Baratta}}, \bibinfo {author} {\bibfnamefont {J.~P.}\ \bibnamefont {Dean}}, \bibinfo {author} {\bibfnamefont {J.~S.}\ \bibnamefont {Dokken}}, \bibinfo {author} {\bibfnamefont {M.}~\bibnamefont {Habera}}, \bibinfo {author} {\bibfnamefont {J.}~\bibnamefont {Hale}}, \bibinfo {author} {\bibfnamefont {C.~N.}\ \bibnamefont {Richardson}}, \bibinfo {author} {\bibfnamefont {M.~E.}\ \bibnamefont {Rognes}}, \bibinfo {author} {\bibfnamefont {M.~W.}\ \bibnamefont {Scroggs}}, \bibinfo {author} {\bibfnamefont {N.}~\bibnamefont {Sime}},\ and\ \bibinfo {author} {\bibfnamefont {G.~N.}\ \bibnamefont {Wells}},\ }\href@noop {} {\  (\bibinfo {year} {2023})}\BibitemShut {NoStop}%
\bibitem [{\citenamefont {Scroggs}\ \emph {et~al.}(2022)\citenamefont {Scroggs}, \citenamefont {Baratta}, \citenamefont {Richardson},\ and\ \citenamefont {Wells}}]{Sscroggs2022basix}%
  \BibitemOpen
  \bibfield  {author} {\bibinfo {author} {\bibfnamefont {M.~W.}\ \bibnamefont {Scroggs}}, \bibinfo {author} {\bibfnamefont {I.~A.}\ \bibnamefont {Baratta}}, \bibinfo {author} {\bibfnamefont {C.~N.}\ \bibnamefont {Richardson}},\ and\ \bibinfo {author} {\bibfnamefont {G.~N.}\ \bibnamefont {Wells}},\ }\href {https://doi.org/10.21105/joss.03982} {\bibfield  {journal} {\bibinfo  {journal} {J. Open Source Softw.}\ }\textbf {\bibinfo {volume} {7}},\ \bibinfo {pages} {3982} (\bibinfo {year} {2022})}\BibitemShut {NoStop}%
\bibitem [{\citenamefont {Aln{\ae}s}\ \emph {et~al.}(2014)\citenamefont {Aln{\ae}s}, \citenamefont {Logg}, \citenamefont {{\O}lgaard}, \citenamefont {Rognes},\ and\ \citenamefont {Wells}}]{Salnaes2014unified}%
  \BibitemOpen
  \bibfield  {author} {\bibinfo {author} {\bibfnamefont {M.~S.}\ \bibnamefont {Aln{\ae}s}}, \bibinfo {author} {\bibfnamefont {A.}~\bibnamefont {Logg}}, \bibinfo {author} {\bibfnamefont {K.~B.}\ \bibnamefont {{\O}lgaard}}, \bibinfo {author} {\bibfnamefont {M.~E.}\ \bibnamefont {Rognes}},\ and\ \bibinfo {author} {\bibfnamefont {G.~N.}\ \bibnamefont {Wells}},\ }\href {https://doi.org/10.1145/2566630} {\bibfield  {journal} {\bibinfo  {journal} {ACM Trans. Math. Softw.}\ }\textbf {\bibinfo {volume} {40}},\ \bibinfo {pages} {1} (\bibinfo {year} {2014})}\BibitemShut {NoStop}%
\bibitem [{\citenamefont {Dalcin}\ \emph {et~al.}(2011)\citenamefont {Dalcin}, \citenamefont {Paz}, \citenamefont {Kler},\ and\ \citenamefont {Cosimo}}]{Sdalcinpazklercosimo2011}%
  \BibitemOpen
  \bibfield  {author} {\bibinfo {author} {\bibfnamefont {L.~D.}\ \bibnamefont {Dalcin}}, \bibinfo {author} {\bibfnamefont {R.~R.}\ \bibnamefont {Paz}}, \bibinfo {author} {\bibfnamefont {P.~A.}\ \bibnamefont {Kler}},\ and\ \bibinfo {author} {\bibfnamefont {A.}~\bibnamefont {Cosimo}},\ }\href {https://doi.org/10.1016/j.advwatres.2011.04.013} {\bibfield  {journal} {\bibinfo  {journal} {Adv. Water Resour.}\ }\textbf {\bibinfo {volume} {34}},\ \bibinfo {pages} {1124 } (\bibinfo {year} {2011})},\ \bibinfo {note} {new Computational Methods and Software Tools}\BibitemShut {NoStop}%
\bibitem [{\citenamefont {Sorre}\ \emph {et~al.}(2009)\citenamefont {Sorre}, \citenamefont {Callan-Jones}, \citenamefont {Manneville}, \citenamefont {Nassoy}, \citenamefont {Joanny}, \citenamefont {Prost}, \citenamefont {Goud},\ and\ \citenamefont {Bassereau}}]{Ssorre2009curvature}%
  \BibitemOpen
  \bibfield  {author} {\bibinfo {author} {\bibfnamefont {B.}~\bibnamefont {Sorre}}, \bibinfo {author} {\bibfnamefont {A.}~\bibnamefont {Callan-Jones}}, \bibinfo {author} {\bibfnamefont {J.~B.}\ \bibnamefont {Manneville}}, \bibinfo {author} {\bibfnamefont {P.}~\bibnamefont {Nassoy}}, \bibinfo {author} {\bibfnamefont {J.~F.}\ \bibnamefont {Joanny}}, \bibinfo {author} {\bibfnamefont {J.}~\bibnamefont {Prost}}, \bibinfo {author} {\bibfnamefont {B.}~\bibnamefont {Goud}},\ and\ \bibinfo {author} {\bibfnamefont {P.}~\bibnamefont {Bassereau}},\ }\href {https://doi.org/10.1073/pnas.0811243106} {\bibfield  {journal} {\bibinfo  {journal} {Proc. Natl. Acad. Sci. U.S.A.}\ }\textbf {\bibinfo {volume} {106}},\ \bibinfo {pages} {5622} (\bibinfo {year} {2009})}\BibitemShut {NoStop}%
\bibitem [{\citenamefont {Aimon}\ \emph {et~al.}(2014)\citenamefont {Aimon}, \citenamefont {Callan-Jones}, \citenamefont {Berthaud}, \citenamefont {Pinot}, \citenamefont {Toombes},\ and\ \citenamefont {Bassereau}}]{Saimon2014membrane}%
  \BibitemOpen
  \bibfield  {author} {\bibinfo {author} {\bibfnamefont {S.}~\bibnamefont {Aimon}}, \bibinfo {author} {\bibfnamefont {A.}~\bibnamefont {Callan-Jones}}, \bibinfo {author} {\bibfnamefont {A.}~\bibnamefont {Berthaud}}, \bibinfo {author} {\bibfnamefont {M.}~\bibnamefont {Pinot}}, \bibinfo {author} {\bibfnamefont {G.~E.~S.}\ \bibnamefont {Toombes}},\ and\ \bibinfo {author} {\bibfnamefont {P.}~\bibnamefont {Bassereau}},\ }\href {https://doi.org/10.1016/j.devcel.2013.12.01} {\bibfield  {journal} {\bibinfo  {journal} {Dev. Cell}\ }\textbf {\bibinfo {volume} {28}},\ \bibinfo {pages} {212} (\bibinfo {year} {2014})}\BibitemShut {NoStop}%
\bibitem [{\citenamefont {Roux}\ \emph {et~al.}(2010)\citenamefont {Roux}, \citenamefont {Koster}, \citenamefont {Lenz}, \citenamefont {Sorre}, \citenamefont {Manneville}, \citenamefont {Nassoy},\ and\ \citenamefont {Bassereau}}]{Sroux2010membrane}%
  \BibitemOpen
  \bibfield  {author} {\bibinfo {author} {\bibfnamefont {A.}~\bibnamefont {Roux}}, \bibinfo {author} {\bibfnamefont {G.}~\bibnamefont {Koster}}, \bibinfo {author} {\bibfnamefont {M.}~\bibnamefont {Lenz}}, \bibinfo {author} {\bibfnamefont {B.}~\bibnamefont {Sorre}}, \bibinfo {author} {\bibfnamefont {J.~B.}\ \bibnamefont {Manneville}}, \bibinfo {author} {\bibfnamefont {P.}~\bibnamefont {Nassoy}},\ and\ \bibinfo {author} {\bibfnamefont {P.}~\bibnamefont {Bassereau}},\ }\href {https://doi.org/10.1073/pnas.0913734107} {\bibfield  {journal} {\bibinfo  {journal} {Proc. Natl. Acad. Sci. U.S.A.}\ }\textbf {\bibinfo {volume} {107}},\ \bibinfo {pages} {4141} (\bibinfo {year} {2010})}\BibitemShut {NoStop}%
\end{thebibliography}
%


\onecolumngrid
\newpage
\begin{center}
\textbf{\large Supplemental Materials for: \\ ``Deformation-induced spectral softening drives phase separation on non-spherical membranes''}
\end{center}
\twocolumngrid
\setcounter{equation}{0}
\setcounter{figure}{0}
\setcounter{table}{0}
\setcounter{section}{0}
\setcounter{page}{1}
\makeatletter
\renewcommand{\theequation}{S\arabic{equation}}
\renewcommand{\thesection}{S-\Roman{section}}
\renewcommand{\thefigure}{S\arabic{figure}}
\renewcommand*{\thepage}{S\arabic{page}}
\renewcommand{\bibnumfmt}[1]{[S#1]}
\renewcommand{\citenumfont}[1]{S#1}

In this Supplemental Material, we provide detailed derivations for the leading order corrections to the Laplace-Beltrami operator in Sec.~\ref{se:LBcorrections}, and the emerging eigenvalues in Sec.~\ref{se:eigenvalues}. We also provide additional details relating the Landau-Ginzburg and the Cahn-Hilliard dynamics in Sec.~\ref{se:LG}, and the numerical procedures used in Sec.~\ref{se:numerics}. Finally, we also provide an argument for the analogue of the experimental sorting ratio $S$ in Sec.~\ref{se:sorting}.

\section{Leading-order correction to the Laplace-Beltrami operator}\label{se:LBcorrections}

We consider a reference surface $\Omega^{(0)}$, characterized by $\bm{X}^{(0)}\left(\bm{u}_{(0)}\right)$, where $\bm{X}^{(0)}\in \mathbb{R}^3$ is the 3-dimensional position vector, and $\bm{u}_{(0)}\!=\!\left(u^1_{(0)},u^2_{(0)}\right)\in\mathbb{R}^2$ is the 2-dimensional surface parameterization. Using this parameterization we define the local tangent vectors $\bm{e}^{(0)}_a\!=\!\frac{\partial \bm{X}^{(0)}}{\partial u^a_{(0)}}$ (for $a\!=\!1,2$), the unit normal  $\bm{n}^{(0)}\!\equiv\!\frac{\bm{e}_1^{(0)}\times\bm{e}_2^{(0)}}{|\bm{e}_1^{(0)}\times\bm{e}_2^{(0)}|}$, the metric tensor $g_{ab}^{(0)}\!\equiv\!\bm{e}_a^{(0)}\cdot\bm{e}_b^{(0)}$, and $g^{(0)}\equiv\det \bm{g}^{(0)}$.

The Laplace-Beltrami operator $\LB$ is written in general form as
\begin{equation}\label{eq:LB_SM}
\LB\bullet\equiv\!\frac{1}{\sqrt{g}}\partial_a\left(\sqrt{g} g^{ab} \partial_b\bullet\right) \ .
\end{equation}
Now we perturb the surface using the normal perturbation $\delta h(\bm{u}_{(0)}) \bm{n}^{(0)}$ to produce the surface $\Omega^{(1)}$. Expanding around $\Omega^{(0)}$, we have~\cite{Sdeserno2015fluid}
\begin{equation}
    \begin{aligned}
    \delta g^{ab}&=-2K_{(0)}^{ab} \delta h \ , \\
    \delta g&=2g^{(0)}K^{(0)}\delta h \ .
    \end{aligned}
\end{equation}
 Using these with the perturbative expansion $g^{(1)}\!\simeq\!g^{(0)} + \delta g\!=\!g^{(0)} \left(1+2 K^{(0)} \delta h\right)$ and $g^{ab}_{(1)}\!\simeq\!g^{ab}_{(0)} + \delta g^{ab}\!=\!g_{(0)}^{ab}-2K^{ab}_{(0)}\delta h$ in Eq.~\eqref{eq:LB_SM}, and retaining the leading order in $\delta h$ yields Eq.~\eqref{eq:LB1} in the main text. For the unit sphere, $K^{(0)}\!=\!2$ and $K_{(0)}^{ab}\!=\!g_{(0)}^{ab}$, so the second term vanishes identically and $\LB^{(1)}\!\simeq\!
\left(1-2\delta h\right)\LB^{(0)}$.

\section{Analytical eigenvalue correction on a deformed sphere}\label{se:eigenvalues}
For the deformed sphere, the correction to the $\LB$ operator is
\begin{equation}
\delta\LB^{(1)}\!\equiv\!\LB^{(1)}-\LB^{(0)}=\!-2 \delta h\LB^{(0)} \ . 
\end{equation}
The associated eigenvalue equation for $\LB^{(1)}$ becomes
\begin{equation}\label{eq:evals}
     \left(1-2\delta h\right)\LB^{(0)}\ket{\psi_n}=\left(\Lambda_n^{(0)} + \delta \Lambda_n\right) \ket{\psi_n}
\end{equation}
where the projection is defined using the metric tensor $\braket{\psi_{m}}{\psi_n}\equiv\oint_\Omega \psi_n^* \psi_m \sqrt{g^{(0)}} d\bm{u}$ (on a sphere $\sqrt{g^{(0)}} d\bm{u}\!=\!\sin \theta d\theta d\phi$). Because the spectrum of the unperturbed sphere is degenerate [with bands of eigenvalues $-l(l+1)$ with a degeneracy of $2l+1$ in each band], the leading order $\ket{\psi^{(0)}_{n}}$ is cast as a superposition of all modes within the degenerate band, $\ket{\psi^{(0)}_{n}}\!=\!\sum_i \alpha_{i}\ket{\psi^{(0)}_{n,i}}$, where we denoted $\psi_{n,i}$ as the $i$th mode in the $n$th degenerate band.

Expanding the state $\ket{\psi_{n}}\!\simeq\!\ket{\psi^{(0)}_{n}} + \ket{\delta \psi_{n}}$, and collecting the leading-order terms we have
\begin{equation}
    \LB^{(0)} \ket{\delta \psi_{n}} - 2\delta h\LB^{(0)}\ket{\psi_{n}^{(0)}} = \Lambda_n ^{(0)}\ket{\delta\psi_{n}} + \delta \Lambda_n \ket{\psi_{n}^{(0)}} \ .
\end{equation}
Contracting this with a different eigenfunction within the same degenerate band $\bra{\psi_{n,j}^{(0)}}$, leads to
\begin{equation}
     \delta \Lambda_n \alpha_j = - 2\alpha_i \Lambda_n^{(0)}\bra{\psi_{n,j}^{(0)}}\delta h \ket{\psi_{n,i}^{(0)}} \ ,
\end{equation}
where we used $\bra{\psi^{(0)}_{n,i}}\LB^{(0)}\!=\!\bra{\psi^{(0)}_{n,i}}\Lambda_n^{(0)}$. This implies $\delta \Lambda_n$ are the eigenvalues of the matrix $-2 \Lambda_n^{(0)}\bra{\psi_{n,j}^{(0)}}\delta h \ket{\psi_{n,i}^{(0)}}$ ($\bra{\psi_{n,j}^{(0)}}\delta h \ket{\psi_{n,i}^{(0)}}$ is evaluated using the $\bm{g}^{(0)}$ metric). 
 
For $\Omega^{(0)}\!=\!\mathbb{S}^2$, $\psi_{n,i}^{(0)}$ correspond to the spherical harmonics $Y_{l,m}$. This implies that to get the analytical predictions for the leading-order corrections, we have to compute the value of $\bra{Y_{l,m_1}}Y_{l_p,m_p}\ket{Y_{l,m_2}}$, where the contraction is between two states in the same degenerate $l$ band, but the perturbation $Y_{l_p,m_p}$ could involve a different $l_p$ band. We compute this using the Wigner 3-j notation. First, we define
\begin{equation}
    \left(\begin{array}{ccc}
        j_1 & j_2 & j_3 \\
        m_1 & m_2 & m_3
        
    \end{array}\right) \! \equiv\! \frac{\left(-1\right)^{j_1-j_2-m_3}}{\sqrt{2 j_3 +1}} \braket{j_1 \ m_1, j_2 \ m_2}{j_3 \ (-m_3)} \ , 
\end{equation}
and where $\braket{j_1 \ m_1, j_2 \ m_2}{j_3 \ (-m_3)}$ are the Clebsch-Gordan coefficients.

\begin{figure}[b!]
\includegraphics[width=0.48\textwidth]{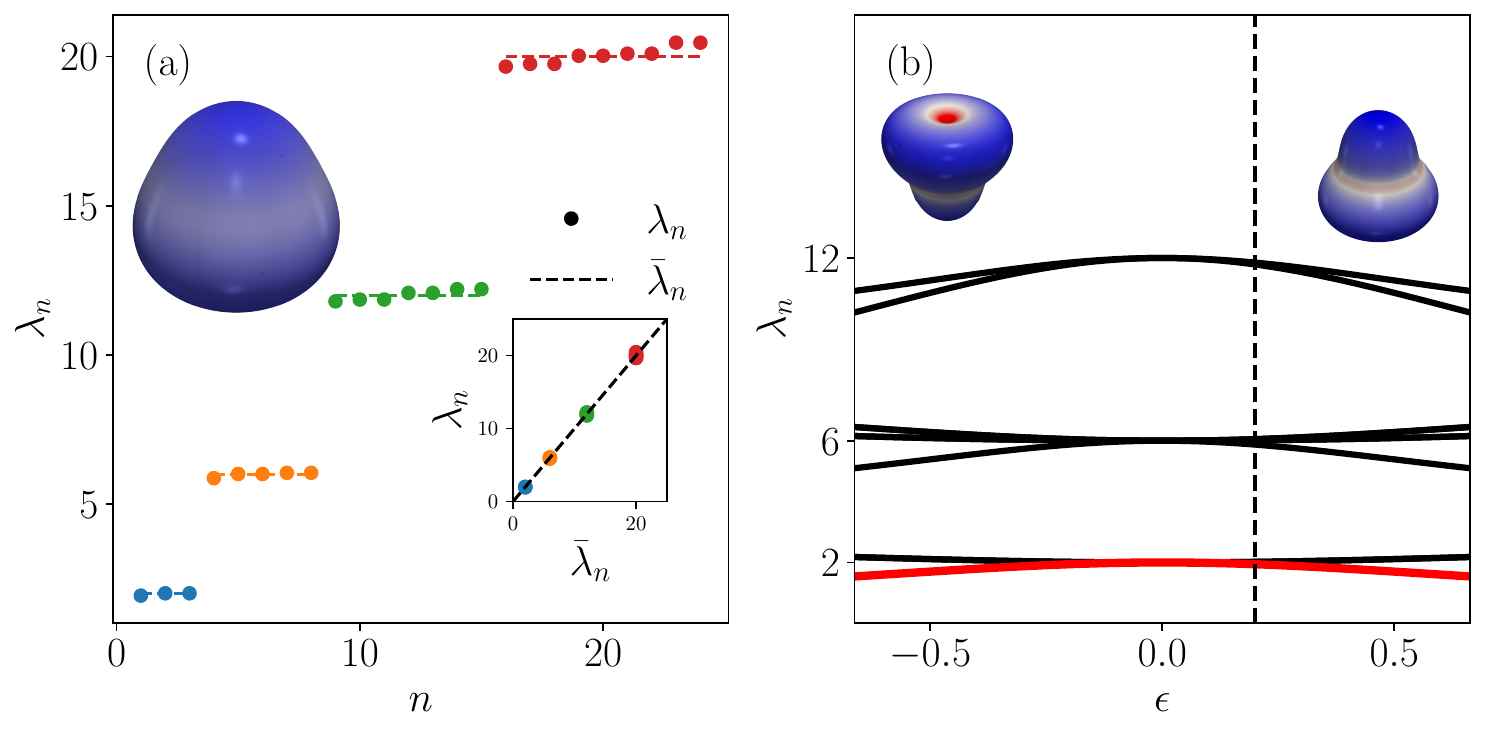}
\caption{Odd $l_p$ values causes the linear corrections to vanish. (a) The numerically computed eigenvalues $\lambda_n$ (points) and the corresponding analytical eigenvalues $\bar{\lambda}_n$ (dashed lines), for the first four spectral bands of a deformed sphere with $\delta h \!=\! 0.2 Y_{3,0}$. Top inset: the deformed sphere (colors represent the local Gaussian curvature). Bottom inset: comparison of the numerical $\lambda_n$ and analytical $\bar{\lambda}_n$ eigenvalues, with the diagonal $\lambda_n\!=\!\bar\lambda_n$ provided as a guide-to-the-eye. (b) The 10 lowest non-zero eigenvalues obtained by deforming the sphere with $\delta h\!=\!\epsilon Y_{3,0}$, as a function of the deformation amplitude $\epsilon$. The lowest non-zero eigenvalue is annotated as a red thick line. The vertical dashed line corresponds to the spectrum shown in (a). Insets: The deformed sphere with $\epsilon\!=\!-0.5$, and with $\epsilon\!=\!0.5$ (left and right, respectively).}
\label{fig:figs1}
\end{figure}

From here we express the elements of the correction matrix $\delta \Lambda_{l,m_1,m_2}$ as
\begin{equation} \label{eq:1st}
\begin{aligned}
    \delta \Lambda_{l,m_1,m_2} \!=\! 2l(l+1)(-1)^{m_1}\sqrt{\frac{(2 l +1)^2 (2 l_p +1)}{4 \pi }}  \\ \times \left(\begin{array}{ccc}
        l & l_p & l \\
        0 & 0 & 0
    \end{array}\right)
    \left(\begin{array}{ccc}
        l & l_p & l \\
        -m_1 & m_p & m_2
    \end{array}\right) \ ,
    \end{aligned}
\end{equation}
where we already substituted $\Lambda_l^{(0)}\!=\!-l(l+1)$. Denoting the $\delta \Lambda_{l,m_1,m_2}$ matrix as $\delta \bm{\Lambda}$, and relating it to $\delta \bm{\Lambda}\!=\!-\delta \bm{\lambda}$, leads to $\lambda^{(1)}\!\simeq\!\lambda^{(0)} + \text{eig}\left[\delta \bm{\lambda}\right]$. This form already implies $\delta \bm{\lambda}\!=\!\bm{0}$ for odd $l_p$ values, due to the first Wigner 3-j term in Eq.~\eqref{eq:1st} (as it requires $2l+l_p$ to be even). We demonstrate this in Fig.~\ref{fig:figs1}, showing the analytical eigenvalue predictions and the numerical calculations in panel (a), and the numerically-computed 10 lowest non-zero eigenvalues in panel (b).

For $m_p\!=\!0$, the correction matrix coefficients are directly computed from Eq.~\eqref{eq:1st}. If we examine $m_p\!\neq\!0$, we add or subtract the associated $\pm m_p$ spherical harmonics to get a real-valued perturbation, e.g., $\frac{1}{\sqrt{2}}\left(Y_{2,2} + Y_{2,-2}\right)$ in the main text. Then the analytical expression is used twice, and we use the sum (or difference) as the correction matrix.

\begin{figure}[b!]
\includegraphics[width=0.48\textwidth]{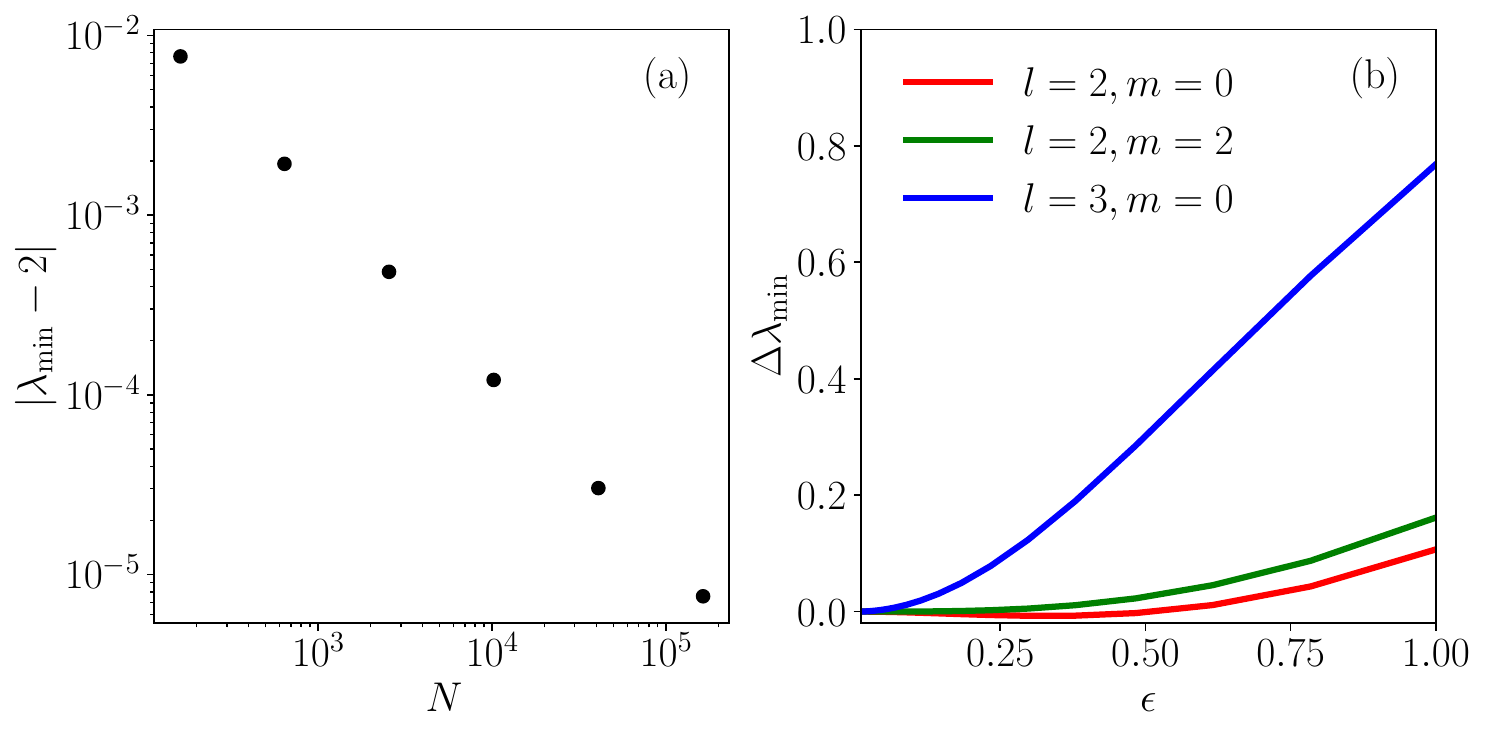}
\caption{Eigenvalue computation errors. (a) The numerical error for $\lambda_{\text{min}}$ for a sphere with increasing mesh points $N$. (b) The error between the numerical $\lambda_{\text{min}}$ and analytical $\bar{\lambda}_{\text{min}}$ values, $\Delta \lambda_{\text{min}}\!\equiv\!\lambda_{\text{min}} - \bar{\lambda}_{\text{min}}$, for different deformation amplitudes $\epsilon$ and different deformation forms (for $l\!=\!3$ we plot $2-\lambda_{\text{min}}$ instead, as the leading order perturbations identically vanish, and because the minimal eigenvalue softens).}
\label{fig:figs2}
\end{figure}

To probe the errors between the analytical and numerical results, we first compare the resulting minimal eigenvalue for icospheres with different resolutions. For $N$ points on the icosphere, we compute $\lambda_{\text{min}}-2$ in Fig.~\ref{fig:figs2}(a), revealing the error decreases as the spatial resolution increases. We use $N\!=\!10242$ for the deformed spheres throughout our work, implying that in the absence of spatial perturbations, the numerical mesh generates an error of $\mathcal{O}\left(10^{-4}\right)$. Then, we probe the difference between the numerical and theoretical minimal eigenvalues $\Delta \lambda_{\text{min}}\!\equiv\!\lambda_{\text{min}} - \bar{\lambda}_{\text{min}}$ as a function of the deformation amplitude $\epsilon$, for the three cases studied above in Fig.~\ref{fig:figs2}(b). For $l\!=\!3$ the first-order corrections vanishes identically (see above), so that $\lambda_{\text{min}}\simeq 2 + \mathcal{O}(\epsilon^2)$ (we plot $2-\lambda_{\text{min}}$ instead, as $\lambda_{\text{min}}$ decreases with $\epsilon$).

\section{Landau-Ginzburg approach and Cahn-Hilliard dynamics}\label{se:LG}
We utilized a free energy functional in Eq.~\eqref{eq:LG} that is similar to the Landau-Ginzburg phenomenological one~\cite{Skardar2007statistical,Ssafran2018statistical}
\begin{equation}\label{eq:freeenergy}
    \beta \mathcal{H}\!=\! \beta F_0 + \int \left[\frac{\tau}{2} \phi^2 + u \phi^4 +\frac{\eta}{2}\left(\nabla\phi\right)^2 - \mu_0 \left(\phi-\phi_0\right)\right] d\Omega \ ,
\end{equation}
where $\beta^{-1}\!\equiv\!k_BT$, $\mathcal{H}$ is the effective Hamiltonian, $F_0$ is the contribution from the short-scale integration, and $\tau$, $u$ and $\eta$ are phenomenological constants that depend on microscopic details and external parameters (e.g., pressure and temperature), and we already set the mechanical energy density $F_\text{m}\!=\!0$. Here we also introduced the Lagrange multiplier $\mu_0$ that enforces $\langle\phi\rangle\!=\!\phi_0$, formally expressing the conservation of $\phi$ on $\Omega$ and absent from Eq.~\eqref{eq:LG} (dynamically, we enforce this conservation using the Cahn-Hilliard dynamics, see below). Using the sphere's radius $R$ as a length-scale in the problem, a dimensionless length ratio that compares the quadratic term to the line-tension would be $\sqrt{\frac{|\tau| R^2}{\eta}}\!\equiv\!\frac{R}{\xi}$, where $\xi$ is the correlation length $\xi\!\equiv\!\sqrt{\frac{\eta}{|\tau|}}$. As we focus here on the role of the underlying geometry, we use the unit sphere $R\!=\!1$, and \emph{a priori} set $\tau\!=\!-1$ (effectively below a critical temperature $T_c$), and $u\!=\!\frac{1}{4}$, implying that $\eta$ sets $\xi$. Fixing the value of $F_0$, this matches the form of Eq.~\eqref{eq:LG}.

\begin{figure*}[t]
\includegraphics[width=0.85\textwidth]{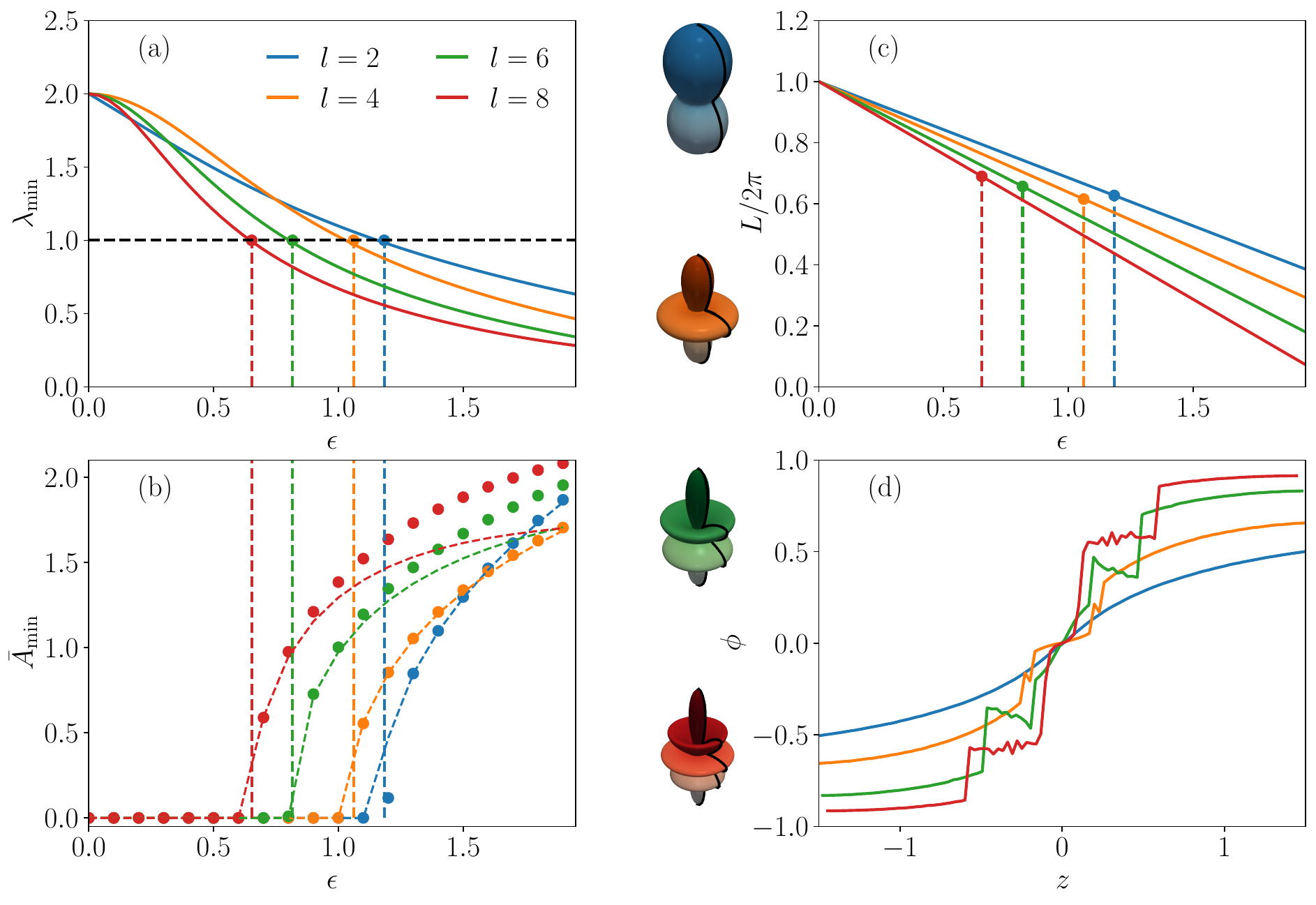}
\caption{Phase separation for different deformations. (a) The eigenvalue $\lambda_{\text{min}}$ evaluated for different deformations $Y_{l,0}$, with increasing amplitude $\epsilon$. Dashed lines correspond to $\lambda_{\text{min}}\!=\!1$. (b) The steady-state $\bar{A}_{\text{min}}$ amplitude (at $t\!=\!100$) as a function of $\epsilon$, for the different deformations. We also plot the theoretical prediction $\sqrt{\frac{1-\lambda_{\text{min}}}{\int \psi_{\text{min}}^4d\Omega}}$ in dashed lines. (c) The normalized minimal azimuthal circumference $L/2\pi$ for the different shapes. (d) The $\phi$ distributions along a $z$-cut at the end of the simulations, for the various $l$ values. Middle insets show the morphologies obtained at $t\!=\!100$ for $\epsilon\!=\!1.5$ for the different $l$ values, showing separated domains (clear versus colored ends). The black curve shows the $z$-cut used in panel (d).}
\label{fig:figs3}
\end{figure*}

Minimizing the energy functional of Eq.~\eqref{eq:freeenergy} $\frac{\delta F}{\delta \phi}\!=\!0$ implies  
\begin{equation} \label{eq:mu}
 \mu_0 = \phi^3-\phi-\eta\nabla^2 \phi \ ,
\end{equation}
identifying with the chemical potential $\mu\!=\!\phi^3-\phi-\eta\nabla^2 \phi$ at steady-state (otherwise $\mu$ could vary spatially). As we assume that the total $\phi$ concentration is conserved, it obeys the continuity equation
\begin{equation}
    \partial_t \phi + \nabla \cdot \bm{j}=0 \ .
\end{equation}
Assuming the flux $\bm{j}\!=\!-M\nabla\mu$, where $M$ is the mobility (taken as $M\!=\!1$)
we arrive at the Cahn-Hilliard equation 
\begin{equation} \label{eq:dynamics}
    \partial_t \phi = \nabla^2 \left[\phi^3-\phi-\eta\nabla^2 \phi\right] \ .
\end{equation}


Perturbing $\phi\!\simeq\!\phi_0 + \delta \phi$ leads to
\begin{equation}
    \partial_{t}\delta\phi=\lambda\left(1-3\phi_{0}^{2}-\eta\lambda\right)\delta\phi \ ,
\end{equation}
implying that once $\eta\lambda+3\phi_{0}^{2}\!<\!1$, the homogeneous $\phi$ state becomes unstable (recall that $\lambda\!\ge\!0$). The first unstable mode then undergoes a linear instability with the rate $\lambda_{\text{min}}(1- 3 \phi_0^2-\eta\lambda_{\text{min}})$, and eventually saturates to a steady-state distribution [shown in the main text in Fig.~\ref{fig:fig3}(a)]. This instability criterion leads to the critical concentration $\phi_c\!=\!\sqrt{\frac{1 - \lambda_{\text{min}}}{3}}$ (for $\eta\!=\!1$), plotted in the $\phi_0-\epsilon$ phase diagram of Fig.~\ref{fig:fig4}(a). For the $\eta-\epsilon$ (with $\phi_0\!=\!0$) phase diagram in Fig.~\ref{fig:fig4}(b) we plot $\eta_c\!=\!1/\lambda_{\text{min}}$ as for this value of $\eta$ the quadratic term and the gradient term exactly balance one another. 

If the \emph{only} unstable mode is the one corresponding to $\lambda_{\text{min}}$, this mode will grow until reaching a steady-state $\partial_t\phi\!=\!0$. This is expected to occur sufficiently close to the instability threshold, where $\lambda_{\min}$ is the only unstable eigenmode and nonlinear coupling to other modes remains weak. Substituting $\phi\!=\!A_{\text{min}} \psi_{\text{min}}$ into Eq.~\eqref{eq:freeenergy}, and minimizing with respect to $A_{\text{min}}$ yields $A_{\text{min}}\!=\!\sqrt{\frac{-\tau-\eta \lambda_{\text{min}}}{4u\int \psi_{\text{min}}^4d\Omega}}$ [shown in Fig.~\ref{fig:fig3}(b) for $\tau\!=\!-1$, $u\!=\!\frac{1}{4}$, and $\eta\!=\!1$].

We demonstrate a similar transition for deformations of different $l$ values in Fig.~\ref{fig:figs3}. We plot $\lambda_{\text{min}}$ for the different $l$ values in Fig.~\ref{fig:figs3}(a), and annotate $\lambda_{\text{min}}\!=\!1$ by dashed vertical lines. The insets at the center show the phase-separated morphologies with their respective colors. Then in panel (b) we show $\bar{A}_{\text{min}}$ for these different deformations, demonstrating that $\bar{A}_{\text{min}}\!\simeq\!\sqrt{\frac{-\tau-\eta \lambda_{\text{min}}}{4u\int \psi_{\text{min}}^4d\Omega}}$ close to the transition $\lambda_{\text{min}}\!=\!1$. At sufficiently large $\epsilon$ values $\bar{A}_{\text{min}}$ deviates from the predicted scaling for $l\!=\!6$ and $l\!=\!8$ --- we suspect this is due to additional modes becoming unstable.

The term $\eta \left(\nabla \phi\right)^2$ sets the interfacial energy (effectively a line tension), implying that at the onset of instability the line tension term becomes preferable, and that the deformation yields a critical length that allows for $\phi$ to phase-separate. The relation between the line tension and $\lambda_\text{min}$ could be understood through the Rayleigh quotient, $\lambda_\text{min}\!=\!\text{min}\frac{\int (\nabla\phi)^2 d\Omega}{\int \phi^2 d\Omega}$~\cite{Scheeger2015lower}.

In Fig.~\ref{fig:figs3}(c) we plot the normalized minimal azimuthal circumference $L/2\pi$ as a function of $\epsilon$, and annotate the critical deformation $\epsilon_c$ to reveal that the interface-length interpretation is qualitatively consistent with the spectral criterion. Finally, we plot in (d) the $\phi$ distribution along a vertical cut through the surfaces (shown as a solid black line on the central insets),  revealing $\sim\!\frac{l}{2}+1$ spatial regions with distinct concentration profiles.

\section{Numerical Simulations}\label{se:numerics}
To simulate the Cahn-Hilliard dynamics on the deformed sphere, we first produced an icosphere mesh~\cite{Ssullivan2019pyvista}. We perturbed the resulting mesh moving the mesh points in their normal direction according to the specific perturbation used~\cite{S2020SciPy-NMeth}, and obtained the eigenspectrum using~\cite{Sreuter2006laplace}.

For the tether numerical simulations, we first define the area of the tether as $2\pi R_T L_T$, where $R_T$ is the tether radius, and $L_T$ is the tether length. We set the tether area to $0.5$. We refine the mesh at the north pole, and pull a tether of radius $R_T$ to length $L_T$ (conserving the area of the tether and of the entire deformed sphere). 

We then initiated a finite-element simulation with mixed elements, of the chemical potential $\mu$ and the composition field $\phi$~\cite{Sbaratta2023dolfinx,Sscroggs2022basix,Salnaes2014unified}. We initialized the $\phi$ field using a uniform distribution between $(-1,1)$, and subtracted the area-weighted average to ensure the distribution fluctuated around zero. We then rescaled these values to have a fixed $L^2$-norm, $\sqrt{\int \phi^2d\Omega}\!=\!0.05$. Finally, we added a constant $\phi_0$ value to shift the distribution with $-0.9\!\le\!\phi_0\!\le\!0.9$, ensuring that $\text{max}|\phi|\!<\!1$.

We then incrementally solved the variational problem using Newton’s method for nonlinear equations~\cite{Sdalcinpazklercosimo2011} with time increments of $\Delta t\!=\!0.01$. Every $100 \Delta t$ we sampled the projected amplitude $A_{\text{min}}$ by projecting the current $\phi$ distribution, on the mode that corresponded to the lowest non-zero eigenvalue, $\psi_{\text{min}}$. 

\begin{figure}[t!]
\includegraphics[width=0.48\textwidth]{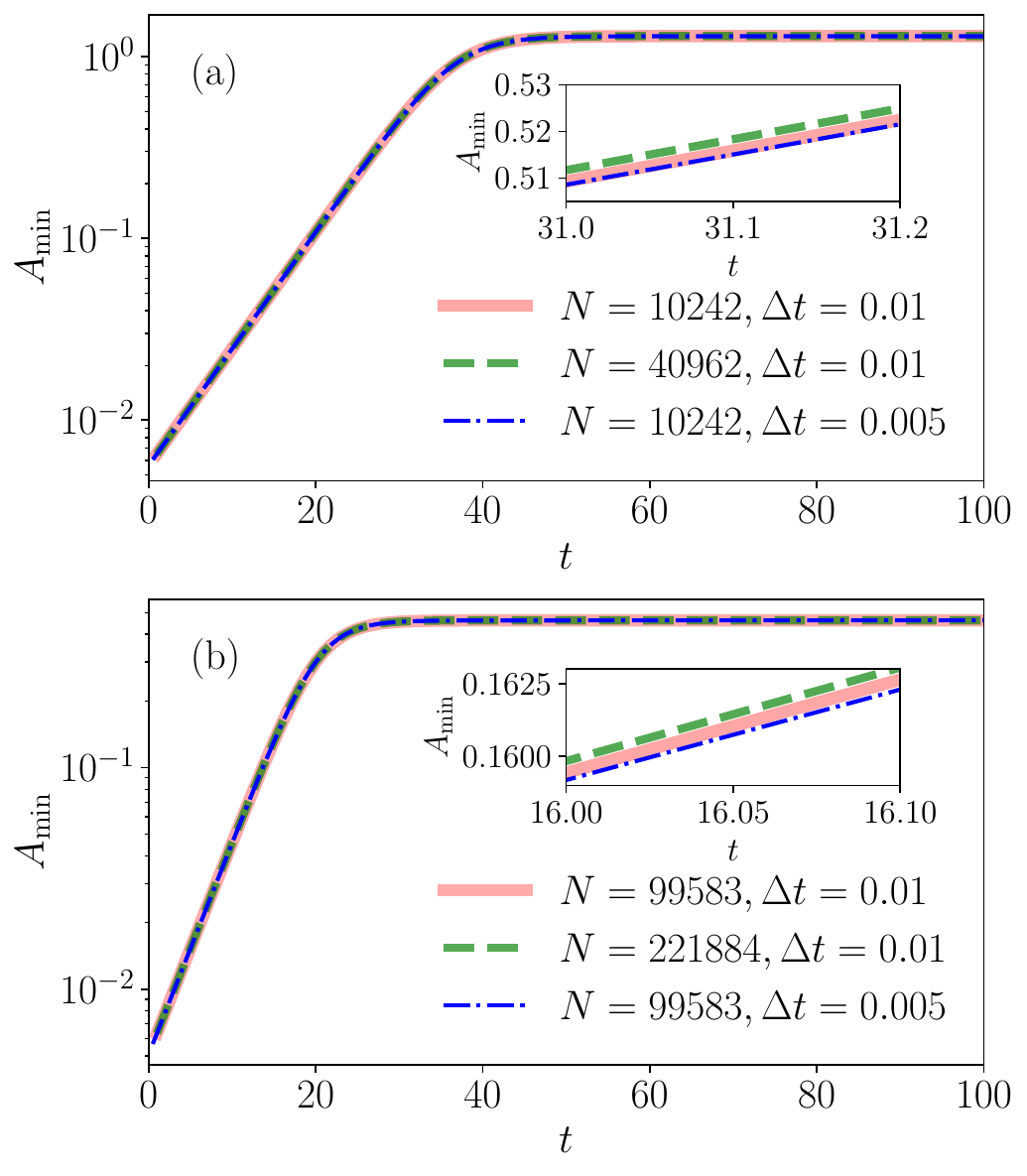}
\caption{Numerical convergence tests for the Cahn-Hilliard dynamics. (a) Semilog plot of the $A_{\text{min}}(t)$ dynamics obtained for $l\!=\!2$, $m\!=\!0$, and $\epsilon\!=\!1.5$, for the spatial and temporal resolutions used throughout our study (red), simulations with increased spatial resolution (green), and simulations with increased time resolution (blue). Inset: a zoom-in to show differences on small scales. (b) Similar convergence test for the tether geometry with $1/R_T\!=\!66.67$ ($R_T\!=\!0.015$).}
\label{fig:figs4}
\end{figure}

Finally, we demonstrate the convergence of the dynamics for two cases in Fig.~\ref{fig:figs4} -- the deformed sphere with $l\!=\!2, m\!=\!0$, and $\epsilon\!=\!1.5$ in (a) and the pulled tether with $R_T\!=\!0.015$ in (b). In both cases, we show the results obtained with the parameters used throughout our work, the results obtained with increased spatial resolution, and the results obtained with increased temporal resolution, revealing numerical convergence.

\section{Experimental sorting ratio}\label{se:sorting}
In many experimental works a central measure for phase separation is the sorting index~\cite{Ssorre2009curvature,Saimon2014membrane,Sroux2010membrane}, defined as the intensity ratio between two markers --- one attached to the membrane denoted here as $I^M$, and another to the constituent of interest (e.g., lipid or protein) denoted here as $I^C$. The ratio $I^C/I^M$ is evaluated at two different geometric areas --- along the tube, denoted here as $I^C_{\text{tube}}/I^M_{\text{tube}}$, and in the rest of the domain, which is usually relatively flat, denoted here as $I^C_{\text{bulk}}/I^M_{\text{bulk}}$. A ratio $I^C/I^M\!=\!1$ indicates the two components scatter light with similar amplitudes, while a value different from $1$ indicates one component scatters more than the other.

Considering the case $\phi_0\!=\!0$, we define the normalized concentration $c\!=\!\frac{1+\phi}{2}$, so that the intensity ratios are proportional to $1+\phi$. The phase-separation mechanism dictates the spatial distribution of $\phi$ but not its relative sign (which depends on the initial conditions). To standardize this, we fix the emerging $\phi$ profile so that $\phi$ at the tip is always positive, and $\phi$ in the bulk is always negative. This is equivalent to considering the sorting ratio to be proportional to 
\begin{equation}
S\!\propto\!\frac{I^C_{\text{tube}}/I^M_{\text{tube}}}{I^C_{\text{bulk}}/I^M_{\text{bulk}}}\!\propto\!\frac{1 + |\phi_{\text{tube}}|}{1 - |\phi_{\text{bulk}}|} \ .
\end{equation}

%

\end{document}